%% file: main.tex
\documentclass[10pt,journal,compsoc]{IEEEtran}

\usepackage{cite}
\usepackage{amsmath,amssymb,amsfonts}
\usepackage{algorithm}
\usepackage[noend]{algpseudocode}
\algrenewcommand\algorithmicindent{0.4cm}
\usepackage{graphicx}
\usepackage{textcomp}
\usepackage{xcolor}
\usepackage{booktabs}
\algnewcommand{\LeftComment}[1]{\State \(\triangleright\) #1}
\algnewcommand{\LeftCommentG}[1]{\Statex \(\triangleright\) #1}
\algnewcommand{\LeftCommentL}[1]{\(\triangleright\) #1}
\algnewcommand\algorithmicforeach{\textbf{for each}}
\algdef{S}[FOR]{ForEach}[1]{\algorithmicforeach\ #1\ \algorithmicdo}
\algrenewcommand\textproc{}
\usepackage{lineno}
\usepackage[breaklinks]{hyperref}
\usepackage{breakurl}
\modulolinenumbers[5]

\usepackage{mwe}  
\usepackage{grffile}
\usepackage{hyperref}
\usepackage{url}
\usepackage{graphicx}
\DeclareGraphicsExtensions{.eps,.pdf}
\usepackage{multirow}
\usepackage{enumerate}
\usepackage{standalone}
\usepackage{tikz}
\usetikzlibrary{calc,backgrounds,arrows,trees,shapes,snakes,shadows,patterns,matrix,fit,patterns,positioning,automata,decorations.pathmorphing,decorations.pathreplacing}
\usepackage{epstopdf}
\usepackage{multicol}
\usepackage{setspace}
\usepackage[aboveskip=-4pt,belowskip=-7pt]{subcaption}
\usepackage[font=footnotesize]{caption} 
\newcommand{\mathtext}[1]{\mathrm{\textit{#1}}}

\newcommand*\lsclass{\tikz[baseline=(char.base),scale=0.7,transform shape]{
    \node[shape=circle,fill=green!50!black,text=white,draw=black,inner sep=2pt] (char) {1};}}

\newcommand*\seclass{\tikz[baseline=(char.base),scale=0.7,transform shape]{
    \node[shape=circle,fill=yellow,text=black,draw=black,inner sep=2pt] (char) {2};}}

\newcommand*\stclass{\tikz[baseline=(char.base),scale=0.7,transform shape]{
    \node[shape=circle,fill=red!80!black,text=white,draw=black,inner sep=2pt] (char) {3};}}

\newcommand*\trone{\tikz[baseline=(char.base),scale=0.7,transform shape]{
    \node[shape=circle,fill=white,text=black,draw=black,inner sep=2pt] (char) {T1};}}

\newcommand*\trtwo{\tikz[baseline=(char.base),scale=0.7,transform shape]{
    \node[shape=circle,fill=gray!30,text=black,draw=black,inner sep=2pt] (char) {T2};}}

\newcommand*\trthree{\tikz[baseline=(char.base),scale=0.7,transform shape]{
    \node[shape=circle,fill=black,text=white,draw=gray!20,inner sep=2pt] (char) {T3};}}

\begin{document}

\title{LFOC+: A Fair OS-level Cache-Clustering Policy for Commodity Multicore Systems}
\author{\IEEEauthorblockN{Juan~Carlos~Saez,~Fernando~Castro,~Graziano~Fanizzi~and~Manuel~Prieto-Matias
\IEEEcompsocitemizethanks{\IEEEcompsocthanksitem Facultad de Inform\'atica, Complutense University of Madrid, Madrid, Spain. 
Email: \{jcsaezal,fcastror,gfanizzi,mpmatias\}@ucm.es.}}}

\IEEEtitleabstractindextext{
\input{abstract}

}

\maketitle

\input{introduction}
\input{related-work}

\input{motivation}

\input{design}

\input{experiments}

\input{conclusions}

\section*{Acknowledgments}

This work has been supported by the EU (FEDER), the Spanish MINECO and CM, under grants RTI2018-093684-B-I00 and S2018/TCS-4423.
\bibliographystyle{IEEEtran}
\vspace{-0.2cm}
\begin{spacing}{0.97}
\bibliography{references}

\vspace{-0.65cm}
\begin{IEEEbiography}[{\includegraphics[height=1.3in,clip,keepaspectratio]{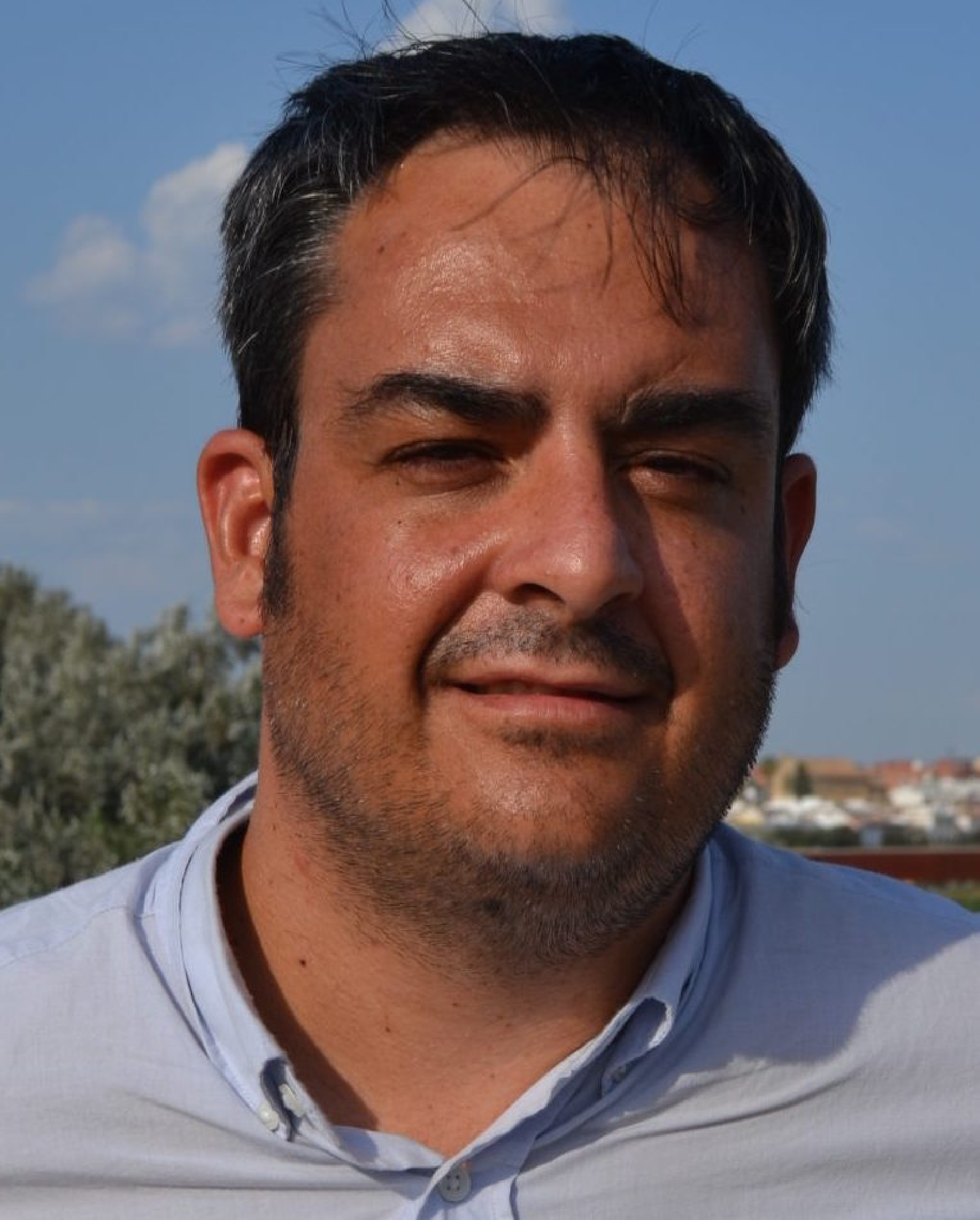}}]{Juan Carlos Saez}received his Ph.D. in computer science from the Complutense University of Madrid (UCM) in 2011. He is now an associate professor in the Dept. of Computer Architecture, and the Campus Representative at UCM of the USENIX international association. His research interests include  system software, scheduling, runtime systems, performance monitoring, and resource management. His recent research activities focus on improving the system software support for emerging hardware platforms.
\end{IEEEbiography}

\vspace{-0.65cm}
\begin{IEEEbiography}[{\includegraphics[height=1.2in,clip,keepaspectratio]{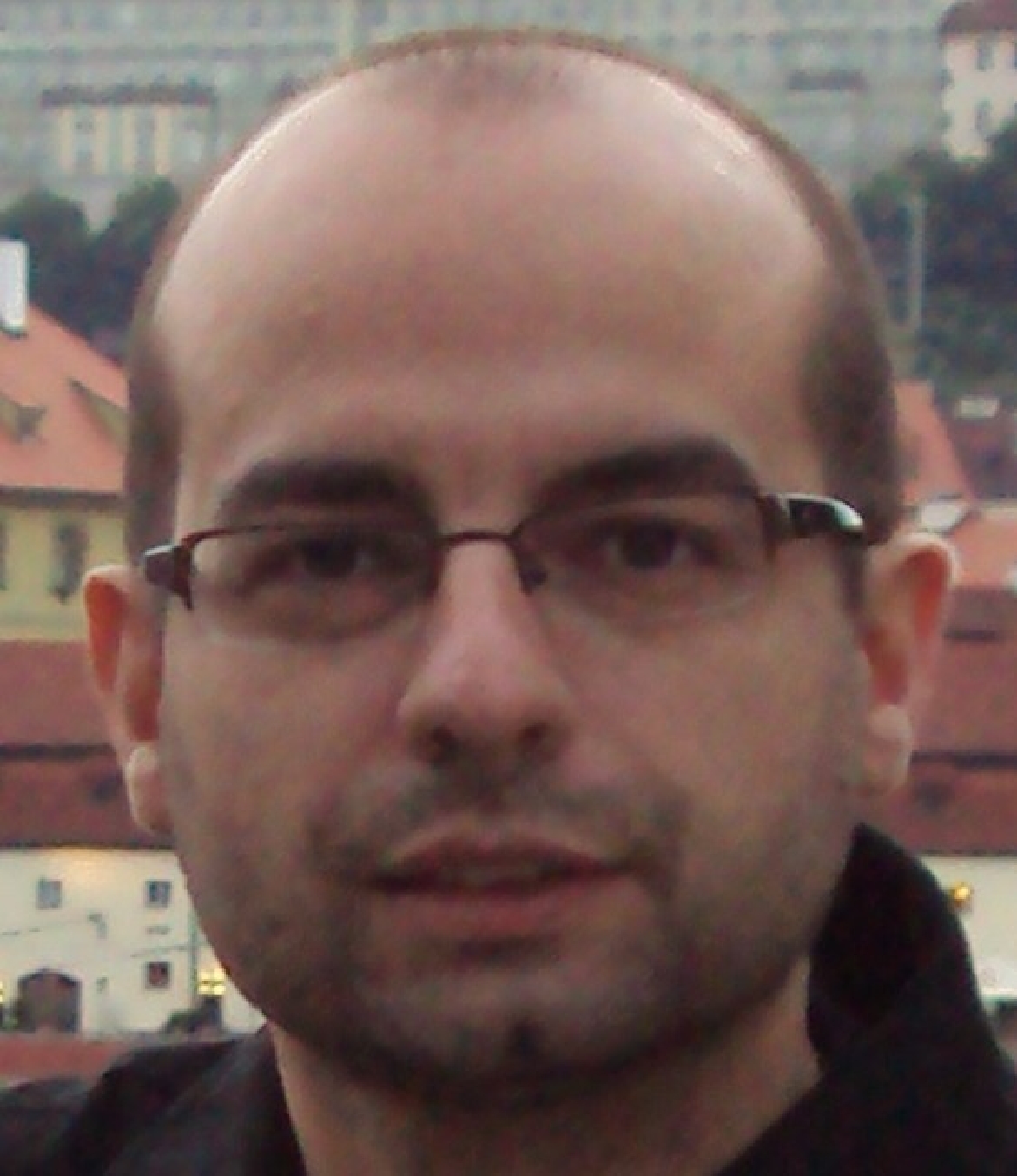}}]{Fernando Castro}received his Ph.D. in computer science from the Complutense University of Madrid (UCM) in 2008. He is now an associate professor in the Dept. of Computer Architecture and his research interests include energy-aware processor design, efficient memory management and OS scheduling on heterogeneous multiprocessors. His recent research activities focus on improving the management of LLCs implemented with emerging technologies.
\end{IEEEbiography}

\vspace{-0.65cm}
\begin{IEEEbiography}[{\includegraphics[height=1.25in,clip,keepaspectratio]{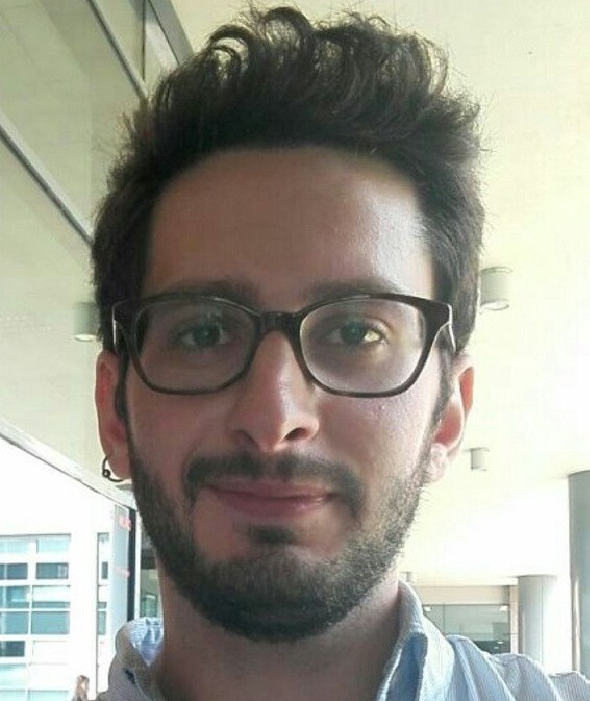}}]{Graziano Fanizzi}received his M.Sc. in Mechatronic Engineering from Politecnico di Torino in 2019. He is now a Ph.D. student in Computer Engineering at Complutense University of Madrid (UCM). His research interests include system software, scheduling and multicore systems. His recent research activities focus on shared resource contention and resource-aware scheduling on multicore systems.
\end{IEEEbiography}

\vspace{-0.65cm}
\begin{IEEEbiography}[{\includegraphics[height=1.25in,clip,keepaspectratio]{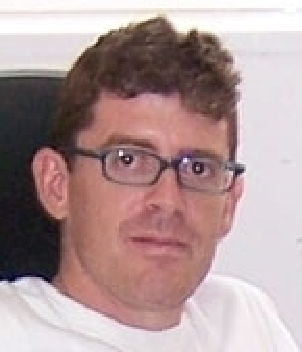}}]{Manuel Prieto}received the PhD degree from the Complutense University of Madrid (UCM) in 2000. He is now full professor in the Dept. of Computer Architecture at UCM.  His research interests include parallel computing and computer architecture.  His current research addresses emerging issues related to heterogeneous systems and energy-aware computing, with a special emphasis on the interaction between the OS and the underlying architecture. He has co-written numerous articles in journals and international conferences on parallel computing and computer architecture.
\end{IEEEbiography}

\end{spacing}

\end{document}

%% file: abstract.tex
\begin{abstract}

Commodity multicore systems are increasingly adopting hardware support that enables the system software to partition the last-level cache (LLC). This support makes it possible for the operating system (OS) or the Virtual Machine Monitor (VMM) to mitigate shared-resource contention effects on multicores by assigning different co-running applications to various cache partitions. Recently cache-clustering (or partition-sharing) strategies have emerged as a way to improve system throughput and fairness on new platforms with cache-partitioning support. As opposed to strict cache-partitioning, which allocates separate cache partitions to each application, cache-clustering allows partitions to be shared by a group of applications.

In this article we propose LFOC+, a fairness-aware OS-level cache-clustering policy for commodity multicore systems. LFOC+ tries to mimic the behavior of the optimal cache-clustering solution for fairness, which we could obtain for different workload scenarios by using a simulation tool. Our dynamic cache-clustering strategy continuously gathers data from performance monitoring counters to classify applications at runtime based on the degree of cache sensitivity and contentiousness, and effectively separates cache-sensitive applications from aggressor programs to improve fairness, while providing acceptable system throughput.

We implemented LFOC+ in the Linux kernel and evaluated it on a real system featuring an Intel Skylake processor, where we compare its effectiveness to that of four previously proposed cache-clustering policies. Our experimental analysis reveals that LFOC+ constitutes a lightweight OS-level policy and improves fairness relative to two other state-of-the-art fairness-aware strategies --Dunn and LFOC--, by up to 22\% and up to 20.6\%, respectively, and by 9\% and 4.9\% on average.
\end{abstract}

%% file: introduction.tex
\section{Introduction}\label{sec:intro}

Chip multicore processors (CMPs) currently constitute the architecture of choice for most general-purpose computing systems, and they will likely continue to be dominant in the near future. Despite the advances in technology, which have made it possible to pack an increasing number of cores and bigger caches on the same chip, contention on shared resources on CMPs still poses a big challenge to the system software. Because cores in a CMP typically share a last-level cache (LLC) and other memory-related resources with the remaining cores --such as a DRAM controller and a memory bus or interconnection network~\cite{ebrahimi10,camps}--, applications running simultaneously on the system may intensively compete with each other for these shared resources, leading to substantial performance degradation~\cite{selfa-pact17}.

Shared-resource contention may introduce a number of undesirable effects on the system, making it difficult to enforce system-wide fairness. For example, contention may cause an application's completion time to differ significantly across runs, depending on its co-runners in the workload~\cite{survey-contention,valencianos-tocs}. In addition, due to contention effects, equal-priority applications may not experience the same performance degradation when running together relative to the performance observed when each application runs alone on the CMP~\cite{ebrahimi10,sergeyb-edp}. These issues make it difficult to provide performance guarantees~\cite{heechul-transc16} or prioritize critical applications without degrading throughput~\cite{ebrahimi10}, may limit scalability of parallel applications~\cite{camps}, and may also cause wrong billings in commercial cloud-like computing services~\cite{valencianos-tocs}.

Previous research has highlighted that shared-resource contention effects can be mitigated by effectively partitioning the shared LLC (i.e., dividing the available cache space among applications). After years of research on cache-partitioning strategies, the necessary hardware support to adopt many of these strategies is now available on commodity processors from Intel (via Cache Allocation Technology - CAT~\cite{cat}), and AMD (as part of QoS Extensions~\cite{amd-rdt}). On these platforms, which  allow the creation of coarse-grained cache partitions only and of a somewhat limited number of partitions, \textit{cache-clustering} (aka \textit{partition-sharing}) algorithms have proven more effective than strict cache partitioning policies~\cite{kpart,pbbcache}. Cache-clustering constitutes a generalization of strict cache partitioning, where, instead of assigning applications to separate cache partitions, each partition can be shared by a group (or cluster) of applications~\cite{icpp-15,pbbcache}. 

Our work explores how to efficiently leverage OS-level cache-clustering to improve fairness on commodity multicores. Our research stands in contrast with recent work on cache-partitioning, which has proposed mostly user-level partitioning approaches that deliver fairness~\cite{selfa-pact17}, or pursue other objectives, such as system throughput optimization~\cite{kpart,cpa-tpds20} or improving client satisfaction on virtual environments~\cite{ginseng-atc16}. In this article we build on our prior work~\cite{lfoc} to advance the state-of-the-art in fairness-aware cache-clustering. Specifically, we propose a new dynamic partitioning approach, referred to as LFOC+, that substantially improves the degree of fairness delivered by our previous proposal --LFOC (Lightweight Fairness-Oriented Cache-clustering); LFOC+ allows cache-sensitive applications (i.e., those that suffer significantly from using reduced cache space) to potentially share the same partition with others, while effectively isolating them from aggressor applications. Our paper makes the following main contributions: \vspace{-0.15cm}

\begin{itemize}
\item Via extensive simulation, we detect that the main limitations of LFOC come from always assigning cache-sensitive programs to separate LLC partitions.
\item To provide a better support for a wider workload range, we design a lightweight cache-clustering algorithm that effectively maps (when beneficial) up to 2 cache-sensitive applications to the same partition.
\item Based on our proposed cache-clustering algorithm, we design and implement LFOC+, an OS-level dynamic partitioning scheme. Our LFOC+'s implementation in the Linux kernel makes efficient use of hardware cache-partitioning extensions at the OS level, and guides cache-clustering by leveraging performance monitoring counters. In implementing LFOC+, we also adapted the OS-level resource-management framework~\cite{lfoc} to be compatible with recent versions of the Linux kernel (v5.x series).
\item We evaluate LFOC+ on a real system featuring an Intel Skylake processor. In our extensive evaluation we qualitatively and quantitatively compare LFOC+ with four state-of-the-art policies: Dunn~\cite{selfa-pact17}, KPart~\cite{kpart}, CPA~\cite{cpa-tpds20} and LFOC~\cite{lfoc}. Our results reveal that LFOC+ improves fairness over these policies for the vast majority of the workload scenarios considered, and operates in a close range of the optimal fairness solution.
 In our analysis we also identify critical design issues of the other approaches, which lead to fairness degradation in some cases. 
\item With respect to our previous work~\cite{lfoc}, we conduct new simulations and experiments, which cover a more ample and diverse set of workloads. We also experiment with data-parallel multithreaded applications, so as to evaluate LFOC+'s unique support to deal with this kind of programs.
\end{itemize}

The remainder of the paper is organized as follows. Section~\ref{sec:related-work} discusses related work. Section~\ref{sec:motivation} presents our extensive simulation analysis, and describes our proposed  cache-clustering algorithm. Section~\ref{sec:design} outlines the design and implementation of LFOC+. Section~\ref{sec:experiments} covers the experimental evaluation, and Section~\ref{sec:conclusions} concludes the paper. 

%% file: related-work.tex
\vspace{-0.35cm}
\section{Related Work}\label{sec:related-work}

A plethora of software and hardware techniques have been proposed to mitigate the effects of contention in the LLC~\cite{survey-contention,survey-cachepart,heracles15,kpart,selfa-pact17,dirigent-asplos16,valencianos-tocs,kim_date19}. Many researchers attempted to address this problem via cache-partitioning~\cite{ucp,cat_first,kpart,whirlpool,dcaps,cpa-tpds20}. A recent survey~\cite{survey-cachepart} discusses a wide range of cache-partitioning approaches for different optimization objectives, such as improving system throughput~\cite{kpart,cpa-tpds20} or fairness~\cite{selfa-pact17,lfoc}. Notably, some recent proposals attempt to enforce QoS constraints for latency critical workloads~\cite{heracles15,ginseng-atc16,parties}.

Cache partitions can be created via specific hardware support (such as Intel CAT) or by means of software solutions, most of which rely on page-coloring~\cite{ics-99,coloring_partitioning_first,pact-14,xiao-eurosys09,taco16}. The various hardware strategies employ different techniques to assign cache ways to different applications; while some of them rely on the cache replacement policy~\cite{khan-hpca14,wang-micro14,quasi-partitioning}, others use set sampling and replicated cache tags~\cite{ucp,subramanian-micro15}. Our proposed OS-level (also extensible to the VMM) strategy leverages hardware-aided way-partitioning.

Recent studies~\cite{pbbcache,lfoc,selfa-pact17} have highlighted that on current CMPs with way-partitioning support, cache-clustering algorithms can be very superior in terms of fairness and throughput than approaches that assign separate partitions to the various applications (aka. strict cache-partitioning). Specifically, the fine-grained distribution of the LLC space that results from sharing a partition among applications, where the space distribution may not be a multiple of the way/set size, could lead to better performance and fairness than what optimal strict cache-partitioning can offer.

UCP~\cite{ucp} is probably the strict cache-partitioning strategy that had a deeper impact on later proposals. UCP relies on \textit{lookahead}, an iterative algorithm that distributes the LLC ways among applications so as to reduce the aggregate number of LLC misses. As indicated in its detailed pseudocode~\cite{ucp}, in each iteration \textit{lookahead} grants a way to the application that experiences the highest reduction in misses when receiving that extra way.  While the original UCP policy used applications' MPKI tables (i.e., Misses per Kilo Instructions for different cache sizes) as input to \textit{lookahead}, recent cache-clustering strategies~\cite{kpart,lfoc} employ variants of the algorithm that are fed with other input metrics, such as the speedup or the slowdown, so as to maximize or minimize the aggregate value of the metric observed across applications. LFOC and LFOC+ use  applications' slowdown tables as input to \textit{lookahead}, used just in part of its clustering-related processing. LFOC is known to deliver better fairness and throughput~\cite{pbbcache} than UCP, and, unlike UCP, it can still be applied when the number of co-running applications exceeds the maximum number of partitions supported by the hardware (this is the case on our experimental setting). 

In this work we compared the effectiveness of LFOC+ to that of LFOC~\cite{lfoc} and Dunn~\cite{selfa-pact17}, which also attempt to improve fairness. The differences between LFOC and LFOC+ are explained in detail in Sec.~\ref{sec:design}. Dunn~\cite{selfa-pact17} relies on grouping applications into clusters, which may overlap, by applying the \textit{k-means} clustering method. In creating clusters, and in determining cluster sizes, Dunn factors in the fraction of stall cycles due to L2 misses for the various applications. As we demonstrate in Sec.~\ref{sec:comparison} this metric can be very misleading to approximate an application's degree of cache sensitivity, and it sometimes leads Dunn to degrade fairness.

We also provide an experimental comparison of LFOC+ against two throughput-optimized cache-clustering policies: KPart and CPA. KPart\cite{kpart} employs an iterative algorithm that creates and merges application clusters via hierarchical clustering. We observed that the distance function used by KPart to decide which clusters to merge, may lead to mapping aggressor and cache-sensitive applications onto the same partition, thus causing substantial fairness degradation in many cases. CPA was proposed later than LFOC~\cite{cpa-tpds20}, but has a few aspects in common with it. Specifically, both LFOC and CPA classify applications into different categories based on their degree of cache-sensitivity and contentiousness, and assign aggressor programs to small partitions. A distinctive feature of LFOC relative to CPA is the fact that the former effectively confines all aggressor (streaming) programs in up to 2 LLC partitions that never overlap with those used for cache-sensitive programs. In addition, LFOC and LFOC+ distribute LLC space between cache-sensitive programs based on their slowdown rather than by considering the number of applications in each class. CPA was originally evaluated using a CMP platform with very different features to those of the one used in our experiments; ours has a different microarchitecture, more cores, bigger L2 caches, coarser-grained LLC partitions, etc. Thanks to the detailed instructions provided by the authors~\cite{cpa-tpds20} to utilize CPA on other LLC configurations, we could evaluate it on our platform. Our evaluation in Sec.~\ref{sec:comparison} reveals that LFOC and LFOC+ provide a higher unfairness reduction than CPA across the board. 

Notably, Dunn, KPart and CPA are user-level clustering approaches, as opposed to LFOC and LFOC+, which were implemented in the OS kernel. User-level solutions require at least an extra user process, and may incur higher overheads due to the use of system calls to access performance monitoring counters (PMCs) and cache partitioning facilities, which are managed by the OS. LFOC and LFOC+ access these facilities directly via a lightweight kernel-level API, and perform PMC-related processing in a distributed fashion (on the CPUs where each thread runs). Another advantage of kernel-level implementations is the fact that they are aware of high-frequency scheduling-related events --like context switches-- and can react to them immediately. This allowed us to efficiently implement LFOC+'s specific support for data-parallel multithreaded applications.

%% file: motivation.tex
\section{Cache-Clustering Algorithms}\label{sec:motivation}

In this section we begin by describing the metrics we used to assess the degree of fairness and throughput of cache-partitioning strategies. Next, we present some details on the simulation tool we used to determine the optimal (fairness-wise) cache-clustering solution for different workloads, and enumerate the different application classes considered in our analysis. Then, we discuss the most relevant patterns detected in the optimal solution, which motivated the design of our earlier LFOC approach~\cite{lfoc}. Finally, we proceed to motivate and describe our newly proposed, more sophisticated cache-clustering algorithm used by LFOC+.

\vspace{-0.2cm}
\subsection{Metrics}\label{subsec:metrics}

To measure the performance degradation of an individual application in a multi-program workload we consider the \textit{Slowdown} metric, defined as follows:
\begin{eqnarray}
\mathtext{Slowdown}_{app}  = \frac{\mathtext{CT}_{\mathtext{part},\mathtext{app}}}{CT_{\mathtext{alone},\mathtext{app}}} \label{eq:slowdown} 
\end{eqnarray}

where \emph{$CT_{\mathtext{part},\mathtext{app}}$} denotes the completion time of application \emph{app} when it runs sharing the system under a given cache-partitioning scheme, and $CT_{\mathtext{alone},\mathtext{app}}$ is the completion time of the application when running alone on the system.

Previous research on fairness for multicore systems~\cite{ebrahimi10,selfa-pact17} defines a scheme as fair if equal-priority applications in a workload suffer the same slowdown as a result of sharing the system. To cope with this notion of fairness, we employ the \textit{unfairness} metric, which has been extensively used in previous work~\cite{ebrahimi10,valencianos-tocs,camps}. For an $n$-application workload, this metric (lower-is-better) is defined as follows: 

\begin{eqnarray}
\textstyle \mathtext{Unfairness} = \frac{\mathtext{MAX}{(}{Slowdown}_{1}{ ,... ,Slowdown}_{n}{)}}{\mathtext{MIN}{(}{Slowdown}_{1} ,... ,{Slowdown}_{n}{)}}  \label{eq:unfairness} 
\end{eqnarray}

To better capture the overall impact of a cache-partitioning approach, the value of the unfairness metric should be reported along with system throughput figures. To quantify throughput, we used the STP metric~\cite{expissue1,selfa-pact17}: 
\begin{eqnarray}
\mathtext{STP} = \sum_{i=1}^{n}\left(\frac{\mathtext{CT}_{alone,i}}{\mathtext{CT}_{part,i}}\right) = \sum_{i=1}^{n}\left(\frac{1}{\mathtext{Slowdown}_{i}}\right) \label{eq:stp}
\end{eqnarray}

\vspace{-0.35cm}
\subsection{Simulation Tool and Application Classes}\label{subsec:simulation}

To determine the optimal (fairness-wise) cache-clustering solution for different workload scenarios, we used the parallel algorithm implemented in the PBBCache simulator~\cite{pbbcache}. This simulation tool has the ability to approximate the degree of throughput, fairness and other relevant metrics for a workload under a particular partitioning algorithm, by leveraging offline-collected application performance data (e.g., instructions per cycle, LLC miss rate, etc.) obtained on the target platform for different cache sizes. For the accurate estimation of an application's slowdown --necessary to determine the unfairness and the STP-- PBB\-Cache accounts for the performance degradation due to both cache sharing and memory-bandwidth contention~\cite{pbbcache}. In assessing the impact of cache sharing, the simulator employs a variant of the Whirlpool method~\cite{kpart,whirlpool} that utilizes applications' LLCMPKC (LLC Misses Per Kilo Cycles) curves to determine how much cache space each application gets when sharing a cache partition with others. This method~\cite{pbbcache} is also used in the implementation of our new LFOC+ proposal.

To obtain the necessary input data for our simulations, we used PMCs to gather the average value of different runtime metrics 
for applications from the SPEC CPU2006 and CPU2017 suites running alone on a real system featuring an Intel Skylake processor with an 11-way 27.5MB LLC (more information on this platform can be found in Sec.~\ref{sec:experiments}). The offline-collected metric values, which correspond to the execution of the first 150 billion instructions of the benchmarks with different cache sizes, are used as input to the simulator. This information is used to determine the optimal solution, namely, the solution to the optimal cache-clustering problem~\cite{pbbcache} that obtains the optimal (minimal) unfairness value for the maximum throughput (STP) attainable.

\begin{figure}[tbp]
 \begin{minipage}{0.49\linewidth}
 \vspace{-0.18cm}
  \centering
   \includegraphics[width=0.96\textwidth]{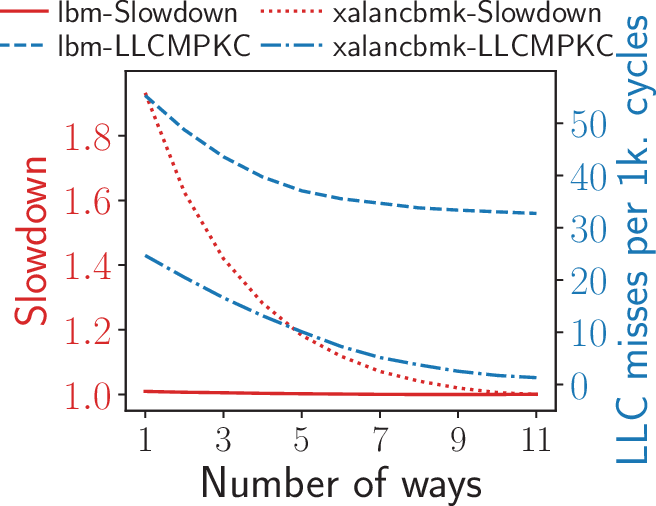}
\vspace{-0.25cm}
  \caption{Slowdown and LLCMPKC for different way counts\label{fig:osci}}
 \end{minipage}
 \hspace{0.1cm}
 \begin{minipage}{0.49\linewidth}
 \vspace{-0.18cm}
  \centering
  \includegraphics[width=0.97\textwidth]{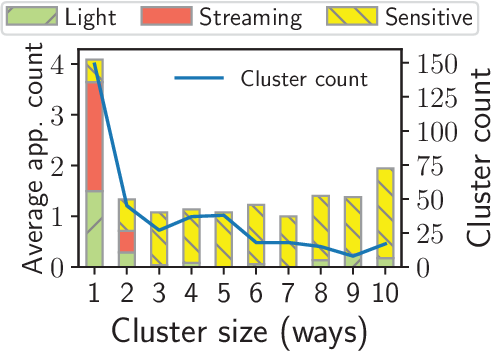}
  \vspace{-0.25cm}
  \caption{Cluster count and breakdown of applications into different categories for each cluster size\label{fig:break}}
 \end{minipage}
\end{figure}

Using the offline-collected data we classified applications into three classes according to their degree of cache sensitivity and contentiousness: \emph{Cache-sensitive}, \emph{light sharing} and \emph{streaming} programs. The \textit{cache-sensitive} category is used for those programs that experience high performance drops as we reduce the number of LLC ways allotted to them; this is not the case for \emph{light sharing} and \textit{streaming} applications. \textit{Streaming} programs are characterized by exhibiting a low slowdown for almost all way allocations, while incurring a high number of LLC misses per cycle. These applications typically have a low cache reuse rate, and degrade the performance of cache-sensitive applications co-located on the same partition~\cite{lfoc}. As an example, Fig.~\ref{fig:osci} shows how the slowdown and the LLCMPKC vary with the number of assigned LLC ways for a streaming application (\texttt{lbm}) and a cache-sensitive one (\texttt{xalancbmk}). \textit{Light-sharing} programs are neither cache sensitive nor aggressive to others (the working set typically fits in the core's private cache levels). 

\subsection{LFOC's Cache-Partitioning Algorithm}

To illustrate the motivation behind our earlier LFOC proposal~\cite{lfoc}, we analyzed the optimal cache-clustering solutions obtained for 114 randomly built workloads (with 4 to 10 applications each). These mixes combine a varying number of streaming, light-sharing and cache-sensitive programs from SPEC CPU2006 and CPU2017. To summarize the behavior of the optimal, Fig.~\ref{fig:break} reports the average application count per cluster size, as well as the total number of clusters --grouped by its size-- that the solution builds.

After a thorough analysis of the solutions obtained, we drew the following three major insights. First, the optimal cache-clustering for fairness always isolates all streaming applications in very small clusters. Specifically, more than 94\% of streaming application instances are assigned to a 2.5MB cluster (1 way), while the remaining ones are allocated to 5MB clusters (2 ways). Second, light-sharing programs are mapped to different clusters following a hardly predictable pattern. Moreover, 90\% of these programs are assigned to 2.5MB clusters. We also observed that moving individual light-sharing programs to other clusters has almost no impact on throughput and fairness. Third, cache-sensitive applications are predominantly present in big clusters: roughly 70\% of these application instances are assigned to clusters with at least 3 ways (7.5MB). This underscores that, in optimizing fairness (i.e., minimizing slowdowns) special care must be taken to fulfill the cache-space requirements of cache-sensitive programs. Moreover, in confining cache-insensitive applications in small clusters, most of the LLC space can be devoted to cache-sensitive programs.

\begin{algorithm}[t]
\caption{Cache-clustering algorithm used by LFOC} \label{alg:clustering}  
\begin{algorithmic}[1]   
\fontsize{9}{8.5}\selectfont
\State {\textbf{Input}: $ST$, $CS$, and $LS$ represent the sets of streaming (str), cache-sensitive and light-sharing applications, respectively; $max\_str\_parts$, $gaps\_per\_str$, and $ways\_str$ are configurable parameters (default values 5 and 3 and 1 respectively
), $nr\_ways$ is the number of ways of the LLC.}
\Statex
\Function{LFOC\_partitioning}{$ST$,$~CS$,$~LS$,$~nr\_ways$}
  \If{$|CS| == 0$}
    \State Create a single cluster $S$ consisting of $nr\_ways$;
    \State Map all applications in $ST \cup LS$ to $S$;
    \State \textbf{return} $\{S\}$
   \EndIf
  \State $Clusters \leftarrow{} \varnothing$; $StreamingClusters \leftarrow{} \varnothing$;
  \LeftCommentG{Step 1: Create as many streaming clusters as needed}  \label{stream-begin} 
  \If{$|ST|>0$}
  \State $parts4str\leftarrow{}\mathrm{\textit{min}}(2,\lceil\frac{|ST|}{max\_str\_parts}\rceil)$; 
  \State $\mathrm{<}r,used{>}\;\;\leftarrow{}\mathrm{<}\lceil\frac{|ST|}{parts4str}\rceil,parts4str*ways\_str\mathrm{>}$;
  \Else
   \State $\mathrm{<}parts4\_str,r,used\mathrm{>}\;\;\leftarrow{}\mathrm{<}0,0,0\mathrm{>}$; 
  \EndIf
  \For{$i \gets 1$ \textbf{to} $parts4str$}
    \State Create new cluster $C$ with $ways\_str$ ways;
    \State Map up to $r$ apps from $ST$ to $C$;
    \State Remove assigned apps from $ST$;
    \State Add $C$ to $Clusters$ and to $StrClusters$;
  \EndFor  \label{stream-end}
  \LeftCommentG{Step 2: Distribute remaining space among apps in $CS$} 
  \LeftCommentG{Use CS apps's slowdown tables as input to lookahead}  \label{sens-begin} 
  \State $W\leftarrow{}$\Call{lookahead}{$CS,\;$nr\_ways-used}; \label{lfocp-line}
  \For{$i \gets 1$ \textbf{to} $|CS|$}
    \State Add a new cluster $C$ with $W[i]$ ways to $Clusters$;
    \State Map application $i$ in $CS$ to $C$;
  \EndFor  \label{sens-end} 
  \LeftCommentG{Step 3: Assign apps in $LS$ to existing clusters}  \label{light-begin}   
  \ForEach{$TargetC \in StreamingClusters$}
    \State $gaps\_avail \leftarrow{} r - |TargetC|*gaps\_per\_str$;
     \If{$|LS|>0$ \textbf{and} $gaps\_avail > 0$}
      \State Map up to $gaps\_avail$ apps from $LS$ to $TargetC$;
      \State Remove assigned apps from $LS$;
     \EndIf 
  \EndFor
  \State {Distribute remaining applications in $LS$ in a round-robin fashion among non-streaming clusters;}  \label{light-end}
  \State \textbf{return} $Clusters$
\EndFunction
\end{algorithmic}
\end{algorithm}

LFOC's cache-clustering algorithm, which is depicted in Alg.~\ref{alg:clustering}, leverages some of the aforementioned insights to enable low-overhead fairness-aware cache partitioning. In Step 1 it reserves up to 2 LLC partitions to map streaming programs. The $ways\_str$ parameter indicates the number of ways used for any of these partitions, which we set to 1 in light of the behavior of the optimal solution in our experimental platform. In Step 2 the remaining LLC space is distributed among cache sensitive applications, which are then assigned to separate partitions. The size of these partitions is determined with \textit{UCP-Slowdown}, namely by applying the \textit{lookahead algorithm} of the UCP policy\cite{ucp} using as input the slowdown curve for each application (i.e., slowdown registered for different cache ways). With this way distribution for cache-sensitive applications LFOC attempts to fulfill their cache requirements based on the degree of cache sensitivity. 
Finally, in Step 3 light-sharing applications are distributed among the various partitions, by populating partitions with streaming applications first.

\vspace{-0.3cm}
\subsection{Fairer LLC Distribution among Sensitive Programs}

Given that LFOC assigns cache-sensitive applications to separate partitions, a question arises as to what extent fairness could be improved by allowing these applications to share the same partition. To provide an answer to this question we conducted additional simulations with 48 randomly built workloads made up exclusively of 6, 8, and 10 cache-sensitive programs. We observed that optimal strict cache-partitioning increases unfairness by 13.7\% on average w.r.t. optimal cache-clustering, and the fairness degradation is higher than 19\% for 25\% of the workloads. Therefore, allowing cache-sensitive applications to share a partition clearly brings substantial fairness improvements.

Unfortunately, efficiently determining how to best group programs into clusters is a big challenge since the number of possible clustering options grows exponentially with the workload size~\cite{pbbcache}. By conducting extra simulations we found that a promising approach is to limit the cluster size to 2. Henceforth, we use the term \textit{Best-2C} to refer to the best fairness-wise cache-clustering for a workload that can be obtained with clusters of up to 2 applications. With \textit{Best-2C} the search-space size is substantially reduced w.r.t optimal cache-clustering, and better unfairness figures can be obtained compared to strict cache-partitioning. Specifically, for 75\% of the 48 workloads the relative fairness degradation of \textit{Best-2C} w.r.t. optimal cache-clustering is now below 3.8\%. This indicates that using a heuristic algorithm that approximates \textit{Best-2C} would bring substantial benefits.

\begin{figure}[tbp]
\vspace{-0.3cm}
\captionsetup[subfigure]{aboveskip=0.5pt,belowskip=0pt}
\centering
\begin{subfigure}{1\linewidth}
\includegraphics[width=1\textwidth]{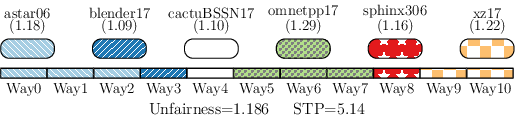}
\caption{UCP-slowdown\label{fig:example-ucps}}
\end{subfigure}
\begin{subfigure}{1\linewidth}
\includegraphics[width=1\textwidth]{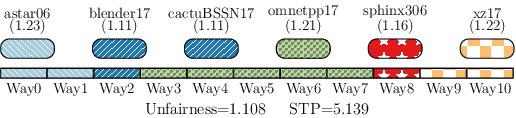}
\caption{Best-2C \label{fig:example-optc2}}
\end{subfigure}
\vspace{-0.2cm}
\caption{Distribution of LLC-space made by (a) UCP-Slowdown and (b) Best fairness-wise clustering with clusters of up to 2 applications\label{fig:example-trans}}
\end{figure}

To guide the design of such an algorithm, we thoroughly analyzed the  differences between the solutions provided by UCP-slowdown (used by LFOC) and \textit{Best-2C} for the 48 workloads. We found that in most cases it is possible to arrive at the second solution from the first one by performing 3 kinds of basic transformations: \trone~transferring one way from a cluster to another, \trtwo~merging two 1-way clusters into a 1-way cluster and transferring the "stolen" way to another cluster, and \trthree~merging two clusters. To illustrate the effectiveness of these transformations let us analyze the example depicted in Fig.~\ref{fig:example-trans}, which displays the solutions of UCP-slowdown and \textit{Best-2C} for a 6-application workload. In the figure, a different color is used to represent each cluster, indicating which cache ways are assigned to it, and which applications make up the cluster (represented by the color shown right below the application name and observed slowdown). In this case, our heuristic algorithm first transfers one way from \texttt{astar06}'s cluster to \texttt{omnetpp17}'s cluster; applying \trone~here enables to reduce the slowdown of \texttt{omnetpp17} (application with the highest slowdown) to 1.25, and in turn the unfairness. To improve fairness even further \trtwo~is applied; \textit{blender17} and \textit{cactuBSSN17} are merged into a single 1-way cluster, and the remaining way is transferred to \texttt{omnetpp17}. This leads to \textit{Best-2C}'s solution.

\begin{algorithm}[tbp]
\caption{Pair-clustering algorithm}\label{alg:pair}
\begin{algorithmic}[1]   
\fontsize{9}{8.25}\selectfont
\State {\textbf{Input}: $CS$ represents the set of cache-sensitive applications, respectively; $S$ represents a matrix that stores the slowdown of each application i for different way counts; $total\_ways$ is the number of cache ways to distribute among applications.} 
\Statex
\Function{pair\_clustering}{$CS,total\_ways,S$}
  
  \State $solutions\leftarrow{}[\;];\;\;\;cost1w\leftarrow{}[\;];\;\;\;\mathtext{cinfo}\leftarrow{}\{\;\}$; 
  \LeftCommentG{Step 1: Determine initial solution (strict partitioning)}
  \State $W\leftarrow{}$\Call{initial\_partitioning}{$CS,\;$total\_ways$,\;S$};
  \State $C\leftarrow{}[\;[app_i]\;\;\mathrm{\bf for}\;\;app_i\;\;\mathrm{\bf in}\;\;CS\;]$;
  \State $\mathrm{<}unf,stp\mathrm{>}\leftarrow{}$\Call{eval\_solution}{$C,W,\;$cinfo};
  \State $solutions.append{(C,W,unf,stp)}$;
  \LeftCommentG{Step 2: Calculate cost of merging any pair of 1-way clusters}
  \For{$i \gets 1$ \textbf{to} $|CS|$}
    \For{$j \gets i+1$ \textbf{to} $|CS|$}

      \If{$W[i]==1$ \textbf{and} $W[j]==1$}

        \State $\mathrm{<}S_i,S_j\mathrm{>}\leftarrow{}$\Call{get\_scurves}{$i,\;j,\;CS,\;S,\;$cinfo};
        \State $cost=\left(S_i[1]+S_j[1]\right)-\left(S[i][1]+S[j][1]\right)$;
        \State $cost1w.append{(cost,i,j)}$;
      \EndIf
    \EndFor
  \EndFor 

  \LeftCommentG{Step 3: Try to apply \trtwo~and~\trthree~to unmerged clusters}
  \State $merged \leftarrow{} [\mathrm{false},\mathrm{false},\cdots ,\mathrm{false}]$\;

  \State $CS_{sorted} \leftarrow{}$ sort applications in $CS$ in descending order   
  \State $\;\;\;\;\;\;\;\;\;\;\;\;\;\;\;\;\;\;\;$by the slowdown for the $W$ assignment

  \For{$app$ \textbf{in} $CS_{sorted}$} \label{loop-improv-begin}
    \State $i\leftarrow{} app.index $;
    \State \textbf{if} $merged[i]$  \textbf{then} \textbf{continue}; 
    \LeftCommentG{Determine the best possible \trtwo~and~\trthree~trans. for app $i$}
    \State \hspace{-0.3cm}$\mathrm{<}cost,a_1,a_2\mathrm{>}\leftarrow{}$\Call{bmerge}{$i,\;$merged$,$~cinfo$,$~cost1w$,\;W,\;CS,\;S$}; 
    \LeftCommentG{If no good transformation found. Skip}
    \If{$cost[i]>0$} \label{good-check} \textbf{continue};  
    \EndIf
    \State $prev\_sol\leftarrow{}C$;
    \If{$a_1\neq{}i$ \textbf{and} $a_2\neq{}i$} 
    \LeftCommentG{Apply~\trtwo: Merge 1-way clusters and move 1 way to app $i$ }
      \State $W[i]\leftarrow{}W[i]+1$; 
      \State Remove entry for $\left(a_1,a_2\right)$ in $cost1w$; 
    \Else
      \ \LeftCommentG{Apply~\trthree: Put $a_1$ and $a_2$ apps in single cluster}
      \State $W[a_1]\leftarrow{}W[a_1]+W[a_2]$; 
    \EndIf 
    \State $merged[a_1]\leftarrow{} true;\;merged[a_2]\leftarrow{}true;\;W[a_2]\leftarrow{}0$;
    \LeftCommentG{Evaluate and store current solution}
    \State $C\leftarrow{}$\Call{build\_new\_solution}{prev\_sol$,\;a_1,\;a_2,\;W$};
    \State $\mathrm{<}unf,stp\mathrm{>}\leftarrow{}$\Call{eval\_solution}{$C,\;W,\;$cinfo};
    \State $solutions.append{(C,W,unf,stp)}$; 
  \EndFor \label{loop-improv-end}

  \State \textbf{return} sol with smallest $(unf,-stp)$ pair in $solutions$
\EndFunction
\end{algorithmic}
\end{algorithm}

\begin{algorithm}[tbp]
\caption{Auxiliary functions used for pair clustering}\label{alg:aux}
\begin{algorithmic}[1]   
\fontsize{9}{8.1}\selectfont

\Function{initial\_partitioning}{$CS,total\_ways,S$}
  \LeftCommentG{The initial solution is provided by \textit{ucp\_slowdown()}}
  \State $W\leftarrow{}$\Call{ucp\_slowdown}{$CS,\;$total\_ways};
  \State $extra\_way \leftarrow{} [\mathrm{false},\mathrm{false},\cdots ,\mathrm{false}]$;

  \For{$it \gets 1$ \textbf{to} $|CS|$} \label{loop-ucpu-begin}
    \State $i\leftarrow{}$ Determine index of application with maximum slowdown for the $W$ assignment;
     \State \textbf{if} $extra\_way[i]$  \textbf{then} \textbf{return} $W$;
    \LeftCommentG{Calculate cost of transferring 1 way to app $i$ (potential~\trone)}
    \For{$j \gets 1$ \textbf{to} $|CS|$}
      \State $\mathrm{<}w_i,w_j\mathrm{>}\leftarrow{}\mathrm{<}W[i],W[j]\mathrm{>}$ ;
      \If{$extra\_way[j]$ \textbf{or} $j==i$ 
       \textbf{or} $w_j==1$ \\ \hspace{2cm} \textbf{or} $S[j][w_j-1]>S[i][w_i+1]$} 
        \State $cost[j] \leftarrow{} \infty $;
      \Else
        \State $cost[j] \leftarrow{} \left(S[i][w_i+1]-S[i][w_i]\right)-\left(S[j][w_j]-S[j][w_j-1]\right)$;
      \EndIf    
    \EndFor
  \LeftCommentG{Perform the best \trone~transformation possible (if any found)}
    \State $<mcost,k>\;\;\leftarrow{}get\_min({cost})$;
    
    \If{$mcost < \infty $} 
      \LeftCommentG{Transfer 1 way: app $k \Rightarrow$ app $i$}
      \State $\mathrm{<}W[i],W[k]\mathrm{>}\leftarrow{}\mathrm{<}W[i]+1,W[j]-1\mathrm{>}$;
      \State $extra\_way[i]\leftarrow{}\mathrm{true}$; 
    \EndIf \label{loop-ucpu-end}
  \EndFor 
\State \textbf{return} $W$
\EndFunction

\Statex
\Function{bmerge}{$i,merged,\mathtext{cinfo},cost1w,W,CS,S$}
  \State $\mathrm{<}bcost,a_1,a_2\mathrm{>}\gets\mathrm{<}\infty,-1,-1\mathrm{>}$;

  \For{$j \gets 1$ \textbf{to} $|CS|$}
    \State \textbf{if} $merged[i]$ \textbf{then} \textbf{continue};
    \State $\mathrm{<}w_i,w_j\mathrm{>}\leftarrow{}\mathrm{<}W[i],W[j]\mathrm{>}$;
    \If{$i==j$} 
    \LeftCommentG{Determine best~\trtwo~and its cost}
      \If{!$cost1w.\mathtext{is\_empty}()$} \label{steal-1-way-begin}
        \State $(cost\_steal,k,l) \gets mincost(cost1w)$;
        \State $cost \gets  S[i][w_i+1]-(cost\_steal)/2$;
        \State $\mathrm{<}a_i,a_j\mathrm{>}\;\;\leftarrow{}<k,l>$;
      \EndIf  \label{steal-1-way-end}
    \Else 
    \LeftCommentG{Determine the cost of merging apps $i$ and $j$~\trthree }
      \State $\mathrm{<}S_i,S_j\mathrm{>}\;\;\leftarrow{}$\Call{get\_scurves}{$i,j,CS,S,\;$cinfo};
      \State $w_c \gets W[i]+W[j]$;
      \State $cost \leftarrow{} \left(S_i[w_c]-S_j[w_c]\right)-\left(S[i][w_i]+S[j][w_j]\right)$;
      \State $\mathrm{<}a_i,a_j\mathrm{>}\;\;\leftarrow{}\mathrm{<}i,j\mathrm{>}\;\;$ \textbf{if} $i\leq{}j\;$\textbf{else}$\;\mathrm{<}j,i\mathrm{>}$;
    \EndIf
    \If{$cost\leq{}0$ \textbf{and} $cost\leq{}bcost$}
      \State $\mathrm{<}bcost,a_1,a_2\mathrm{>}\;\gets\;\mathrm{<}cost,a_i,a_j\mathrm{>}$
    \EndIf
  \EndFor
    \State \textbf{return} $<bcost,a_1,a_2>$
\EndFunction
\end{algorithmic}
\end{algorithm}

Algorithm~\ref{alg:pair} depicts our \textit{pair-clustering} heuristic method, which aproximates \textit{Best-2C}. This new algorithm is used by LFOC+ to distribute LLC space among cache-sensitive applications. Its first step is to invoke \texttt{initial\_partitioning()} --defined in Alg.~\ref{alg:aux}--, which provides an initial cache-space distribution that employs strict cache-partitioning. This function retrieves the way assignment of UCP-slowdown, and iteratively tries to improve fairness by applying~\trone~transformations to the initial way assignment. In each iteration of the loop in lines \ref{loop-ucpu-begin}-\ref{loop-ucpu-end} of Alg.~\ref{alg:aux}, it finds the partition of the application with the highest slowdown --$i$-- and determines the best possible~\trone~transformation that grants an extra way to application $i$: the transformation that provides the highest reduction of its slowdown without increasing too much the slowdown of the application that yields a way.

Before performing additional transformations to the initial solution, Step 2 of Alg.~\ref{alg:pair} calculates the cost of merging any pair of 1-way clusters in the initial solution into a 1-way cluster, so as to guide~\trtwo~transformations. Step 3 (lines \ref{loop-improv-begin}-\ref{loop-improv-end}) traverses applications in the initial solution in descending order by slowdown, and successively applies~\trtwo~or~\trthree~transformations to reduce the slowdown of the current application. If the partition has not already been merged, the \textit{bmerge()} auxiliary function (defined in Alg.~\ref{alg:aux}) is invoked. This function calculates the cost of the ~\trtwo~and~\trthree~transformations that are more beneficial to the current application, and selects the best transformation (i.e., the one that provides the highest slowdown reduction without making the aggregate slowdown grow). Back in Alg.~\ref{alg:pair} (line \ref{good-check}), if a good transformation was found, the best type of transformation (~\trtwo~or~\trthree) is performed giving rise to a new solution, which is evaluated and stored in the \texttt{solutions} list. The algorithm returns the cache-clustering solution with the smallest unfairness value in the list; if two or more solutions with the same unfairness exist, the one with the highest throughput (STP) value is selected.

Finally, we describe the purpose of the \texttt{build\_new\_solution()} and \texttt{get\_scurves()} auxiliary functions used in Alg.~\ref{alg:pair}. The first function returns a new solution that can be obtained by merging --in the existing \textit{prev\_sol} solution-- the two single-application clusters whose IDs are passed as a parameter ($a_1$ and $a_2$), and subject to the distribution of cache ways in $W$. The \texttt{get\_scurves()} function returns two slowdown curves, which hold an estimation of the slowdown that applications $i$ and $j$ experience when assigned to the same partition, and for different partition sizes. To obtain these slowdown curves we used the same Whirlpool-based method employed by PBBCache~\cite{pbbcache}. Notably, to avoid redundant calculations, once the curves for two specific applications have been calculated, they are stored in the \texttt{cinfo} data structure. So, in later function invocations for the same two applications, the curves are directly retrieved from \texttt{cinfo}.

%% file: design.tex
\vspace{-0.2cm}
\section{The LFOC+ dynamic partitioning strategy}\label{sec:design}

In Sec.~\ref{sec:lfoc-design} we describe the main differences between the design and implementation of the LFOC and LFOC+ dynamic partitioning strategies. Next, in Sec.~\ref{sec:other-features} we showcase how LFOC+ deals with multithreaded applications as well as with the contention that may arise due to confining streaming programs in small cache partitions.

\vspace{-0.2cm}
\subsection{Design and Implementation}\label{sec:lfoc-design}

We have implemented LFOC and LFOC+ using a loadable kernel module in Linux v5.4.55.  Specifically, the implementation has been bundled as a \textit{monitoring plugin} (extension of the OS scheduler) in the PMCTrack tool~\cite{pmctrack-compj}, which features a kernel-level API to access privileged hardware facilities such as PMCs and cache-partitioning extensions.

LFOC and LFOC+ continuously monitor the value of several PMC metrics for the various applications to classify them at runtime into the \textit{light sharing}, \textit{streaming} and \textit{sensitive} classes. Both schemes have two operating modes: \textit{sampling} and \textit{fairness}. The \textit{sampling mode} is activated every so often to determine an application's class. During the \textit{fairness} mode, they execute their corresponding partitioning algorithm periodically.  In LFOC, Algorithm~\ref{alg:clustering} is used for LLC-space distribution; in LFOC+ a variant of that algorithm is used, where the only difference lies in the LLC-space distribution method for cache-sensitive programs (line \ref{lfocp-line} of Alg.~\ref{alg:clustering}). Specifically, LFOC uses \textit{UCP-Slowdown} for this task, whereas LFOC+ leverages the \textit{pair-clustering} strategy (Alg.~\ref{alg:pair}).

When a new application (process) enters the system its cache behavior is unknown, so an \textit{unknown} class is initially assigned to it. Right after its creation, each thread has to go through a warm-up period (3 sampling intervals in our setting). PMC data gathered during the warm-up period is not used to classify applications, so as to mitigate mispredictions associated with cold-start effects (e.g., spikes in the LLC miss rate may be present at the start of the execution). A \textit{sampling list} is maintained to keep track of the applications whose warm-up period has elapsed, and those for which a transition between cache-sensitivity classes was detected. This list is checked periodically during the \textit{fairness} mode. When it is not empty, the application found at the list head is removed, and a transition is triggered into the \textit{sampling mode}, so as to determine the application's current class.

\begin{figure}
\vspace{-0.2cm}
 \begin{minipage}[tbp]{0.5\textwidth}
  \begin{minipage}[tbp]{0.33\textwidth}
  \centering
  \includegraphics[width=1.01\textwidth]{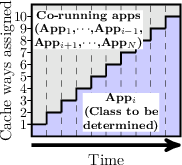}
  \vspace{-0.6cm}
  \captionof{figure}{Sampling mode's LLC space distribution\label{fig:lfoc-sampling}}
  \end{minipage}
  \hspace{0.1cm}
\begin{minipage}[tbp]{0.67\textwidth}
  \fontsize{8.5}{8.25}\selectfont
  \addtolength{\tabcolsep}{-3pt}    
  \begin{tabular}{cl}
    \bf Type & \bf Criterion\\
    Streaming & ($Slowdown\leq{}1.03$ \\ 
              &  and $LLCMPKC\geq{}10$)\\  
              & in at least 1 way assignment, \\ 
              & and $Slowdown < 1.06$ \\
              & in  all  way assignments\\
   Sensitive & If not streaming \\
            & and $Slowdown\geq{}1.05$ \\ 
          & for a number of ways $\geq 2$ \\
  Light shar. & otherwise 
    \end{tabular}
      \vspace{-0.25cm}
      \captionof{table}{Classification of applications} 
      \label{tab:classif}
    \end{minipage}
  \end{minipage}
\end{figure}

The sampling mode --depicted in Fig.~\ref{fig:lfoc-sampling} for the 11-way LLC of our experimental platform-- is inspired by the technique proposed by El-Sayed et al.~\cite{kpart}. Two non-overlapping cache partitions covering the entire LLC are created; the first one, referred to as the \textit{sampling partition}, is reserved for the application that triggered the transition into sampling mode, and the other one is devoted to the remaining applications. As we gradually increase the sampling-partition size, and the remaining applications begin to receive a smaller amount of LLC space, the value of various PMC events (i.e., instructions retired, processor cycles, LLC misses and LLC occupancy) is gathered so as to later determine the application class. 
Two main differences exist between the earlier technique~\cite{kpart} --for the KPart strategy-- and ours. First, in the former approach the size of the sampling application is progressively smaller (rather than bigger). Second, unlike the former method, ours does not require a full sweep of all LLC ways, but can be interrupted after exploring a few way counts, which contributes to substantially reduce the overhead that may come from the sampling mode. Notably, this complete sweep is necessary for KPart~\cite{kpart}, as its partitioning algorithm receives as input the MPKI and IPC values of each application for all possible cache ways. In LFOC and LFOC+, the sampling process is interrupted as soon as making the sampling partition bigger provides no useful information to the clustering algorithm. The heuristic  to cancel the sampling process relies on monitoring the LLCMPKC metric; intuitively, as soon as a very low number of LLC misses is detected when increasing the size of the sampling partition, we expect that performance (IPC) remains very close after assigning more ways, so exploring bigger cache sizes is not necessary. In doing so, light-sharing applications can be rapidly identified: they incur very few LLC misses after exploring only 1 or 2 ways. In a similar vein, many cache-sensitive programs begin to exhibit a light-sharing like behavior (i.e., low number of misses) as soon as its cache-space requirements are fulfilled.

At the end of the sampling period, the application's slowdown curve is built by using the IPC values gathered for different cache sizes, and employing the highest IPC observed as a reference for the calculation (estimate of the performance with all LLC ways). After that, the application class is determined by using a set of rules. Table~\ref{tab:classif} summarizes the classification rules employed on our experimental platform, which are defined in terms of the application slowdown, and the LLCMPKC. We obtained these classification rules after analyzing the value of various PMC metrics gathered offline for 15 distinct applications with different cache sizes (we collected only 50 billion instructions for each application and way count). In designing our classification method, a main goal was to make it simple enough --as~\cite{quasi-partitioning,cpa-tpds20}-- so that it could be evaluated efficiently at the OS level.

Once the application class has been determined, the obtained slowdown curve is stored in the task structure for cache-sensitive programs only (for LFOC+ the LLCMPKC curve is stored as well), as this information is necessary for the partitioning algorithms. If the sampling list is empty, a transition is automatically triggered into the \textit{fairness} mode. When entering this mode, the associated partitioning algorithm is executed immediately, to quickly undo the suboptimal LLC distribution enforced during the sampling mode.

Because an application may exhibit different program phases, its initial classification may not be representative throughout the execution. To reduce potential overheads introduced by periodic sampling-mode activations, this mode is only triggered when the application class has likely changed.  To detect class transitions our implementation leverages lightweight PMC-based mechanisms that strive to react to coarse-grained program phases only, thus minimizing the number of sampling-mode activations. For light-sharing programs, a class change is signaled if it enters a memory-intensive phase, namely, if a running average of the LLCMPKC exceeds a \texttt{high\_threshold} (10 in our platform, as reported in Table~\ref{tab:classif} for streaming-like behavior).  Conversely, for streaming programs, usually assigned to small LLC partitions, the sampling mode is engaged if its average LLCMPKC falls below a  \texttt{low\_threshold} (defined as 30\% of \texttt{high\_threshold}). Finally, for sensitive applications, a heuristic is employed to detect abrupt changes in its slowdown curve without actually rebuilding it. In particular, a \textit{critical size} is associated with each sensitive application. This size --determined during its last sampling period-- indicates the point of the curve (number of ways) where the slowdown falls below 1.05 (as in Table~\ref{tab:classif}).  A class change is signaled when a non-memory intensive stable phase is detected (inverse criterion used for light-sharing applications) for a LLC occupancy smaller than the critical size, or when the average LLCMPKC$>$\texttt{high\_threshold}  for a LLC occupancy  bigger than the critical size. 

We should highlight that LFOC and LFOC+ are resource managers primarily conceived to work in scenarios without oversubscription, just like most recent cache-clustering proposals~\cite{selfa-pact17,kpart,cpa-tpds20}. Nevertheless, to effectively deal with scenarios when the total number of threads exceeds the platform's core count, our implementation could be adapted so as to work in closer cooperation with the OS scheduler. Specifically, our resource-management proposals are more amenable for integration with contention-aware co-schedulers that also rely on application classification~\cite{valencianos-tocs,fairness-griegos}. So, combining fairness-aware cache-clustering with co-scheduling of the most suitable subset of threads would enable to reduce contention even further.

\vspace{-0.2cm}
\subsection{Dealing with Multithreaded Applications and Reducing Contention of Streaming Programs}\label{sec:other-features}

LFOC+ seeks to deliver fairness by grouping applications into clusters and mapping these clusters to LLC partitions of potentially different size. Unlike all previously proposed cache-clustering policies~\cite{selfa-pact17,kpart,cpa-tpds20} --including LFOC~\cite{lfoc}--, LFOC+ is equipped with support for dealing with regular data-parallel multithreaded applications. In these programs, all threads do mostly the same kind of processing with different data, so the values of the various PMC metrics are very similar across threads. To reduce the amount of sampling required for online characterization, LFOC+ selects just one thread --referred to as the \textit{reference} thread-- in each program. During the sampling mode, the PMC metrics gathered for this thread alone are used to determine the application class by following the same method described in Sec.~\ref{sec:lfoc-design}. At all times, all threads in an application are assigned to the same LLC partition by LFOC+. When the application-to-partition mapping is changed by the partitioning algorithm, all threads are immediately migrated to the assigned partition. To this end, our OS-level implementation relies on inter-processor interrupts (IPIs); when the partitioning algorithm completes on a certain core (it changes over time),  LFOC+ issues an IPI-based cross-call for each runnable thread in the application to update the partition ID in a register of the corresponding CPU~\cite{cat}.

Because threads in a multithreaded application may block due to synchronization or other events (e.g., page faults), the application's \textit{reference} thread can vary over time. LFOC+'s implementation tracks thread activations and deactivations during context switches, and ensures that the selected reference thread is always the runnable thread with the smallest value of the \texttt{pid} field in Linux task structure. In OpenMP programs, for example, this strategy selects the \textit{master} as the reference thread most of the time, as it typically spends a larger portion of its execution in the runnable state w.r.t. other threads. In Sec.~\ref{sec:multithread} we assess the effectiveness of this support by using workloads that include data-parallel multithreaded programs. We leave for future work the inclusion of support for other types of multithreaded applications, like those that exploit pipeline parallelism. For these applications, threads can be grouped in separate sets according to their processing type (e.g., based on the \textit{main} function they execute), and a reference thread can be selected within each group so as to efficiently determine the application-wide class.

LFOC+ also takes special care with the contention that may arise due to confining streaming programs in 2.5MB partitions (1-way on our platform). While these programs do not suffer noticeably when running alone on the system with 1 LLC way, we observed that co-running many streaming programs competing for the same way may be harmful due to the huge number of memory requests that result from that competition. This issue may substantially limit the effective memory bandwidth of each streaming program, and, in turn, degrade their performance.
In the workloads we used, this does not have a significant impact in fairness but usually translates into throughput degradation, especially when the number of memory-intensive applications in the workload is high. To effectively mitigate this problem, LFOC+ merges multiple 2.5MB 1-way streaming partitions into a single 5MB (2-way) partition. Likewise, we observed that assigning a streaming-like multithreaded application to a 1-way partition dramatically increases the competition between its various threads, and it may lead to substantial throughput degradation. Therefore, as a conservative measure, LFOC+ also ensures that streaming multithreaded applications are confined (when necessary) in 5MB 2-way partitions. As we demonstrate in Sec.~\ref{sec:multithread}, this provides a better tradeoff between fairness and throughput.

%% file: experiments.tex
\vspace{-0.3cm}
\section{Experiments}\label{sec:experiments}

For our evaluation we used a 20-core server platform featuring a Xeon Gold 6138 ``Skylake'' processor where cores run at 2Ghz. This processor integrates an 11-way 27.5MB LLC (L3) with way-partitioning support; each core features a 64KB L1 and a 1MB L2 cache. On this platform we carried out an experimental comparison of LFOC+ with Stock-Linux (i.e., the vanilla Linux kernel without modifications, which does not partition the LLC), and with the Dunn~\cite{selfa-pact17}, KPart~\cite{kpart}, LFOC~\cite{lfoc} and CPA~\cite{cpa-tpds20} partitioning policies.

\begin{figure*}[t]
\vspace{-0.3cm}
\centering
\includegraphics[width=1\textwidth]{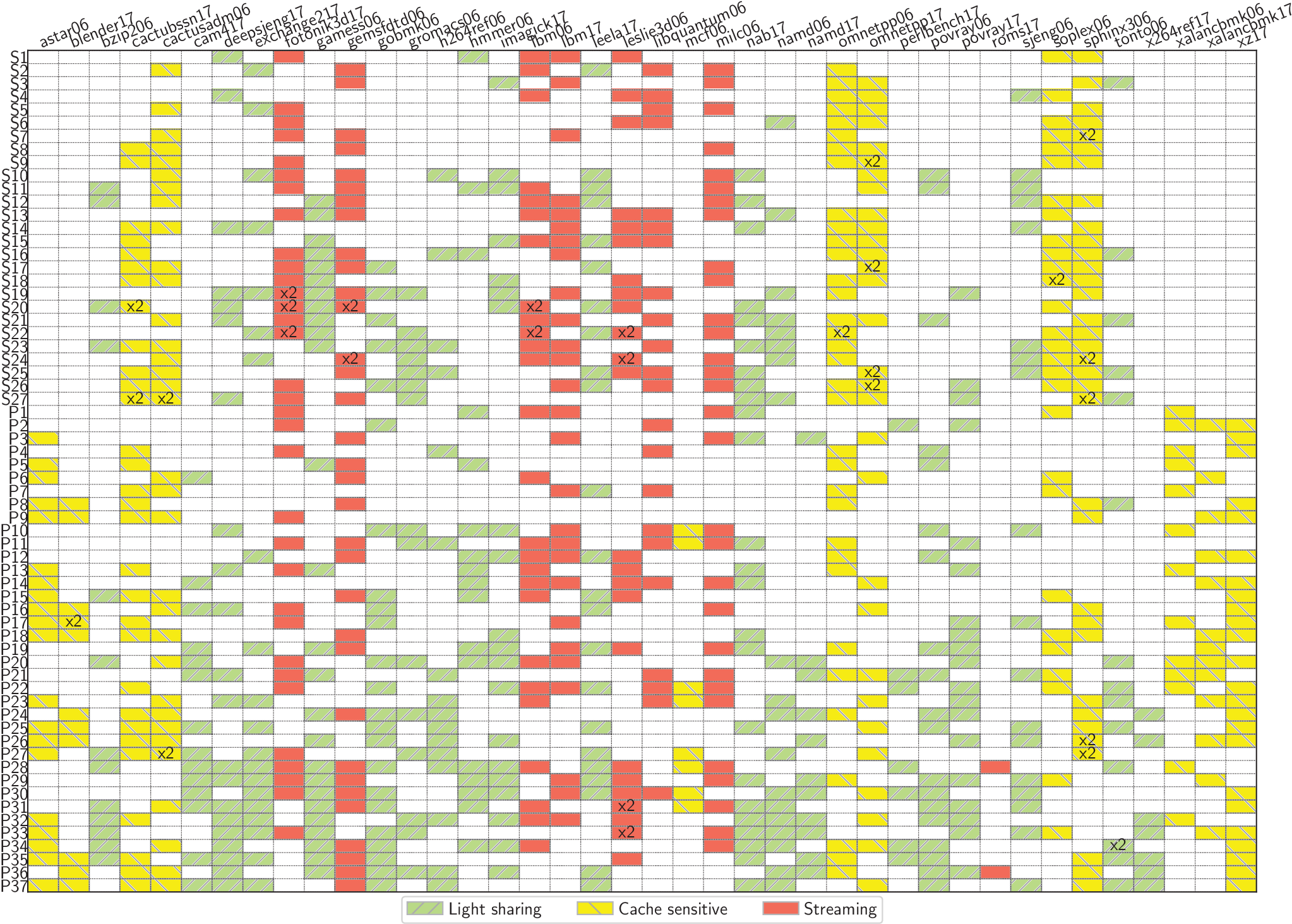}
\vspace{-0.6cm}
\caption{Multiprogram workloads used for our experiments. The ``x2'' mark indicates that 2 instances of a benchmark are present in a workload\label{fig:composition}}
\vspace{-0.4cm}
\end{figure*}

In Sec.~\ref{sec:comparison} we evaluate the effectiveness of LFOC+ when running workloads consisting of single-threaded benchmarks from SPEC CPU using the reference input set. This kind of workloads was employed in the evaluation of all previous approaches~\cite{selfa-pact17,kpart,lfoc,cpa-tpds20}. In Sec.~\ref{sec:overhead} we perform an overhead analysis of LFOC+'s sampling mode. In Sec.~\ref{sec:multithread} we discuss the results provided by LFOC+ when using workloads that include multithreaded applications.  

In conducting multi-application experiments we follow a similar methodology to that of previous work~\cite{expissue2,expissue1,expissue3,camps}, which avoids runs of which significant portions are spent executing only the slowest program(s) in the workload. Specifically, we ensure that all applications in the mix are started simultaneously, and when one of them completes, the program is restarted repeatedly until the longest application in the set completes three times. We then measure unfairness and STP, by using the geometric mean of the completion times for each program. To ensure a fair comparison with previous proposals when using workloads made up of sequential programs (Sec.~\ref{sec:comparison}) we complete only a fixed number of instructions for each application (i.e., 150B in our setting), as most previous proposals~\cite{selfa-pact17,kpart,lfoc} were evaluated by considering only portions of an application's execution. Conversely, for workloads that include multithreaded programs (Sec.~\ref{sec:multithread}), we perform a full execution of all the applications in each mix, as running a fixed instruction count in a multithreaded program does not always capture the same execution portion across runs~\cite{wood06}.

\vspace{-0.35cm}
\subsection{Comparison with Previous Approaches}\label{sec:comparison}

Fig.~\ref{fig:composition} depicts the composition of the 64 randomly generated workloads used for our experiments in this section. These program mixes include varying amounts of streaming, cache-sensitive and light-sharing benchmarks. Note that we selected 40 applications from both the SPEC CPU2006 and CPU2017 suites to experiment with a wider range of streaming and cache-sensitive programs, as most benchmarks in both suites exhibit a light-sharing cache-insensitive execution profile on our platform. This is caused in part due to the coarse granularity of the LLC partitions we can create on this system (the smallest partition is as big as 2.5MB) and the somewhat large private L2 caches (1MB). We used workloads of 8, 12, 16 and 20 applications each, so as to analyze the impact that the workload size has on fairness. Notably, these workloads are more diverse than those explored in our previous work~\cite{lfoc}, which included no more than four cache-sensitive applications each. The vast majority of SPEC benchmarks not present in the workloads of Fig.~\ref{fig:composition} exhibit a light-sharing behavior, which is already represented well by 50\% of the selected programs. For the sake of completeness, workloads in Sec.~\ref{sec:multithread} do include most of the benchmarks not considered here. 

In the remainder of this section we first evaluate the effectiveness of the cache-clustering algorithms used by the various partitioning strategies, when fed with offline-collected PMC data. Then we evaluate how fairness-aware dynamic clustering schemes, which rely on online-collected data, perform when applications exhibit different phases.

\textit{A) Evaluation of Clustering Algorithms.} The main goal in this section is to assess the degree of fairness and throughput delivered by a certain clustering strategy (i.e., how applications are grouped into shared or separate partitions according to their runtime properties) putting aside the associated overheads due to algorithm execution, performance monitoring and cache allocation.

To assess the effectiveness of each clustering algorithm, we consider the first 27 workloads (S$i$ mixes shown in Fig.~\ref{fig:composition}) consisting of 8, 12 or 16 applications whose behavior falls in a clear class (i.e., cache sensitive, streaming or light sharing) for the vast majority of the execution. For the analysis, we implemented all the clustering algorithms used by KPart, CPA, Dunn, LFOC and LFOC+ on top of the PBBCache simulator. To conduct the corresponding experiments, we launched the simulator prior to the execution of each workload to retrieve the cache partitions and application-to-partition mappings imposed by a certain clustering strategy. Then, we enforced the corresponding cache partitions on a per-process manner from user-space, and launched the workload in question, which used the same \textit{static} cache configuration throughout the execution. Note that, while in our simulation analysis of Sec.~\ref{sec:motivation} we employed the Unfairness values provided by PBBcache~\cite{pbbcache}, here we calculate the values of this metric with the actual completion times of the programs executed on the real system.

\def\tamfig{1}

\begin{figure*}[tbp]
\vspace{-0.25cm}
\centering
\includegraphics[width=\tamfig\textwidth]{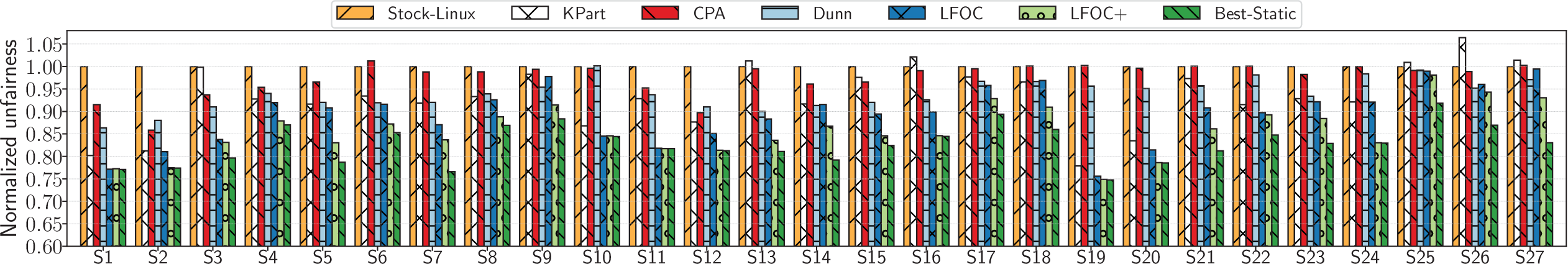}
\includegraphics[width=\tamfig\textwidth]{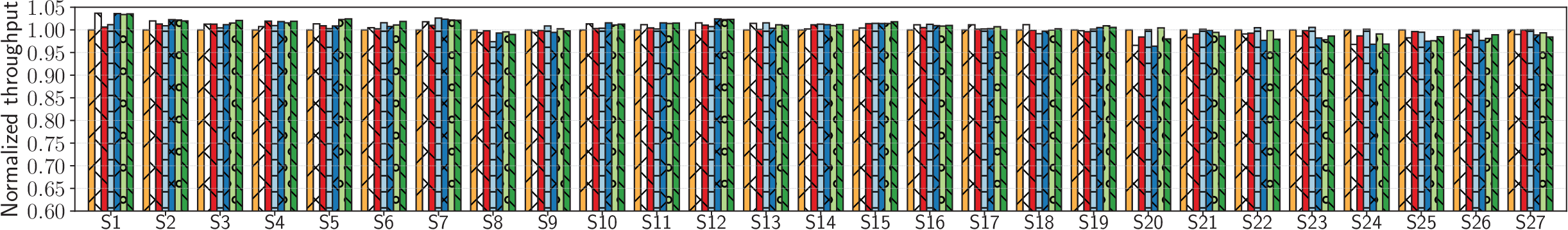}
\vspace{-0.6cm}
\caption{Normalized unfairness and STP values obtained by the static version of the various clustering algorithms\label{fig:static}}
\end{figure*}

\begin{figure}
 \begin{minipage}[tbp]{1\textwidth}
  \begin{minipage}[tbp]{0.28\textwidth}
  \vspace{-0.6cm}
  \centering
\hspace{-0.1cm}
\includegraphics[width=1\textwidth]{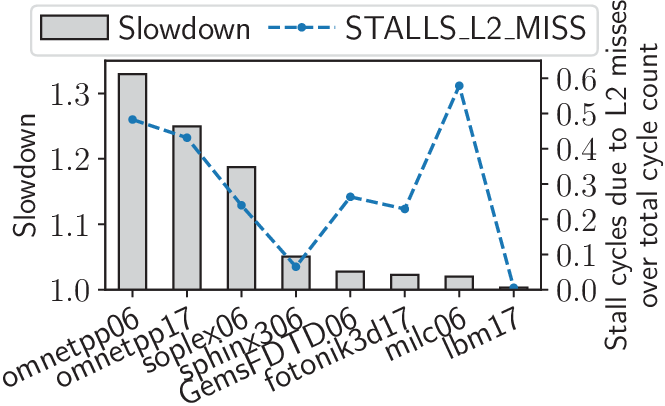}
\vspace{-0.7cm}
 \captionof{figure}{Slowdown vs. stall cycles due to L2 misses for 4 cache-sensitive (left) and 4 streaming (right) programs running alone and using 4 LLC ways\label{fig:correl}}
  \end{minipage}
  \hspace{0.15cm}
\begin{minipage}[tbp]{0.2\textwidth}
\vspace{-0.25cm}
\captionof{table}{Average and max. unfairness reduction of the various strategies relative to Stock-Linux for experiments of Fig.~\ref{fig:static}\label{table:mean}}
\vspace{-0.2cm}
\centering
\footnotesize
\fontsize{8}{8.5}\selectfont
\addtolength{\tabcolsep}{-3pt}
\begin{tikzpicture}[scale=0.94,transform shape,inner sep=0pt]
\node (table1) {\begin{tabular}{rrr}
\toprule
  \textbf{Strategy} &      \textbf{Avg.} &      \textbf{Max.} \\
\midrule
KPart &  6.65\% &  22.12\% \\
CPA &  2.49\% &  14.19\% \\
Dunn &  6.06\% &  13.65\% \\
LFOC & 10.62\% &  24.39\% \\
LFOC+ & 14.21\% &  25.15\% \\
Best-static  & 17.27\% &  25.22\% \\
(optimal)  &  &  \\
\bottomrule
\end{tabular}};
\end{tikzpicture}
    \end{minipage}
  \end{minipage}
\end{figure}

\begin{figure}
\vspace{-0.4cm}
\captionsetup[subfigure]{aboveskip=1pt,belowskip=-6pt}
\centering
\begin{subfigure}{0.46\linewidth}
\includegraphics[width=1\textwidth]{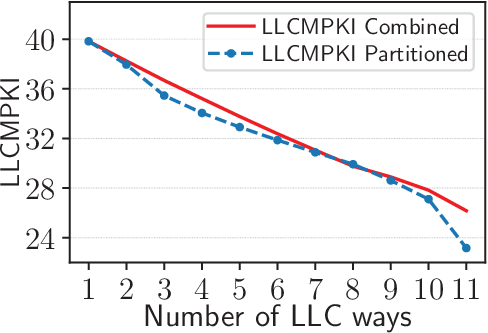}
\caption{Aggregate LLCMPKI\label{fig:llcmpki-curve}}
\end{subfigure}
\begin{subfigure}{0.46\linewidth}
\includegraphics[width=1\textwidth]{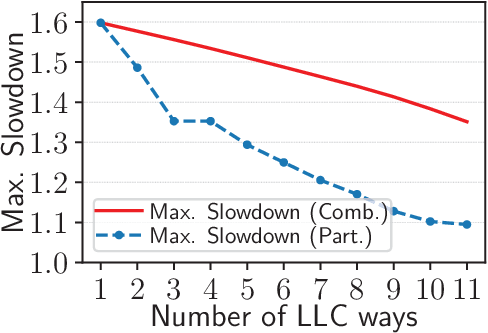}
\caption{Max. Slowdown\label{fig:maxslowdown-curve}}
\end{subfigure}
\caption{Combined and partitioned curves (\texttt{libquantum06}-\texttt{omnetpp17})}
\end{figure}

Fig.~\ref{fig:static} shows the degree of unfairness and throughput delivered by the different clustering strategies; the values have been normalized to the results of Stock-Linux. With this first set of experiments, we also provide a comparison with the optimal fairness cache-clustering solution determined by the PBBCache simulator (see Sec.~\ref{subsec:simulation}), denoted as \textit{Best-Static} in the figure. This comparison with the optimal is an important contribution of our research. As with the other approaches, we used static cache partitioning based on the information provided by PBBCache for the optimal solution. We should highlight that to make this comparison with the optimal feasible we had to use mixes of up to 16 applications, as the exponential growth of the search space in this optimization problem made it impossible to obtain the solution for higher application counts. By leveraging a parallel algorithm~\cite{pbbcache} on our 20-core experimental platform, PBBCache determines the optimal solution for 16-application mixes in roughly 9.5 hours, where 10.48 billion clustering solutions are explored. The 82.8 billion possible clusterings for a 17-application workload, or the 51.7 trillion options to explore for a 20-application workload makes it impractical to determine the optimal in these scenarios. 

KPart's clustering algorithm, designed to optimize throughput, surprisingly brings modest throughput gains (up to 3.67\%). Furthermore, we observe that it sometimes brings substantial fairness degradation over LFOC and LFOC+ (e.g., for the S3, S13 or S26 workloads). We found that KPart fails to systematically map aggressor streaming applications and cache-sensitive programs to separate partitions, which is crucial to reduce unfairness. This behavior is caused by the distance function that KPart uses to decide whether to map several applications to the same partition. Specifically, for S26, KPart opts to assign the \texttt{libquantum06} (streaming) and \texttt{omnetpp17} (highly cache-sensitive) programs to the same partition. Fig.~\ref{fig:llcmpki-curve} shows the partitioned and combined LLCMPKI curves that KPart uses to determine the distance between the two aforementioned applications. These curves depict the aggregate LLCMPKI obtained when assigning the two applications to separate and to the same partition, respectively, for different partition sizes; the closer the curves, the smaller the distance, and hence the more likely for KPart to combine both applications in the same partition (which is the case in S26). However, if we analyze how the maximum slowdown varies when mapping both applications to the same partition (see Fig.~\ref{fig:maxslowdown-curve}), we can conclude that assigning them to separate partitions is clearly more beneficial in terms of fairness for most partition sizes. So, our overarching conclusion is that the KPart's distance function alone is not appropriate to guide fairness-aware cache-clustering decisions, as it relies on the LLCMPKI metric, which is known to be a misleading indicator in the context of shared-resource contention~\cite{contention-sensitivity}.

As KPart, CPA also strives to improve throughput. However, CPA provides very modest throughput improvements relative to Stock-Linux (by up to 2\% for S4) in our experimental platform, whose cache hierarchy differs from that of the system where CPA was originally evaluated~\cite{cpa-tpds20}. The observed fairness gains are smaller than 5\% for over 85\% of the workloads. These small improvements stem from the fact that most of the workloads we explored include more than 3 highly/mildly cache-sensitive programs, referred to as \textit{critical} applications in CPA's classification~\cite{cpa-tpds20}. In these cases, CPA assigns all critical applications to one partition that spans the entire LLC, thus reducing fairness and throughput optimization opportunities. One aspect that contributes to reducing unfairness slightly in most workloads is the fact that CPA confines up to 2 aggressor applications in separate LLC partitions (of 2.5MB in our platform). Programs detected as aggressors in CPA are those that either fall in a special \textit{squanderer} category or non-critical applications that are later detected to have a high LLC occupancy. Unfortunately, when the number of aggressors in the workload exceeds 2, CPA assigns the remaining aggressors to the same partition devoted to non-critical programs. Because this partition may overlap with those reserved to mildly and highly cache-sensitive programs, fairness can be severely degraded,
as our experiments reveal.  By contrast, LFOC+ effectively isolates streaming aggressor programs (such as \texttt{milc06} or \texttt{lbm06}) in small partitions that never overlap with those devoted to cache-sensitive programs, enabling to improve fairness by 12\% on average over CPA.

The results of the Dunn policy --designed to optimize fairness-- exhibit a non-uniform behavior across workloads; in some cases (e.g., S1) Dunn reduces unfairness by up to 13.6\%, but in many others it obtains modest fairness improvements over Stock-Linux. This is due to the fact that Dunn frequently assigns streaming and cache-sensitive applications to the same partition, or to different but overlapping partitions. This stems from its exclusive reliance on the number of processor stall cycles due to L2 cache misses; the higher the value of this event, the higher the number of LLC ways allotted by Dunn to the application~\cite{selfa-pact17}. Fig.~\ref{fig:correl} shows that there is no clear correlation between an application's L2-miss stall cycles and the slowdown it suffers when running with reduced LLC space. For example, in the S10 workload, where Dunn slightly degrades fairness, the \texttt{soplex} (sensitive) and \texttt{GemsFDTD} (streaming) programs are assigned by Dunn to the same partition --due to the similar value of the aforementioned metric-- along with other sensitive programs. Mixing this kind of aggressor applications with highly cache-sensitive programs in the same partition leads to performance degradation and unfairness. So, we conclude that using the L2-miss stall cycles metric alone is not enough to fairly distribute space in the LLC. 

\def\tamff{0.93}

\begin{figure*}[tbp]
\vspace{-0.25cm}
\centering
\includegraphics[width=1\textwidth]{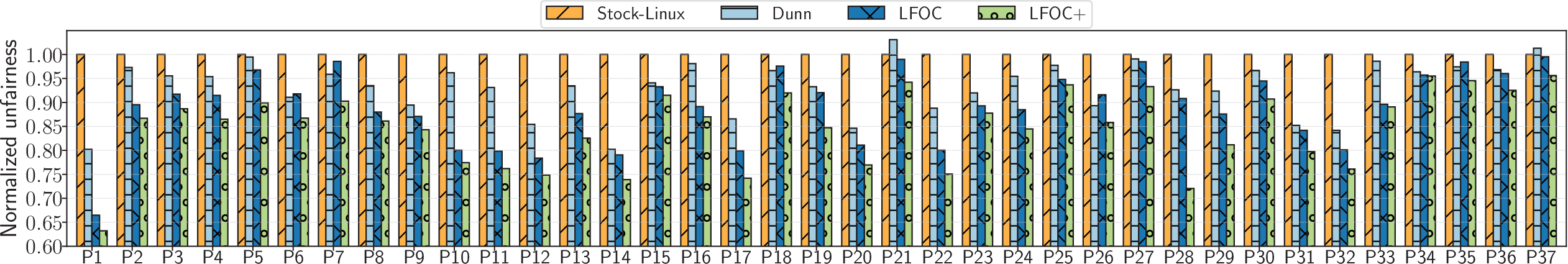}
\includegraphics[width=1\textwidth]{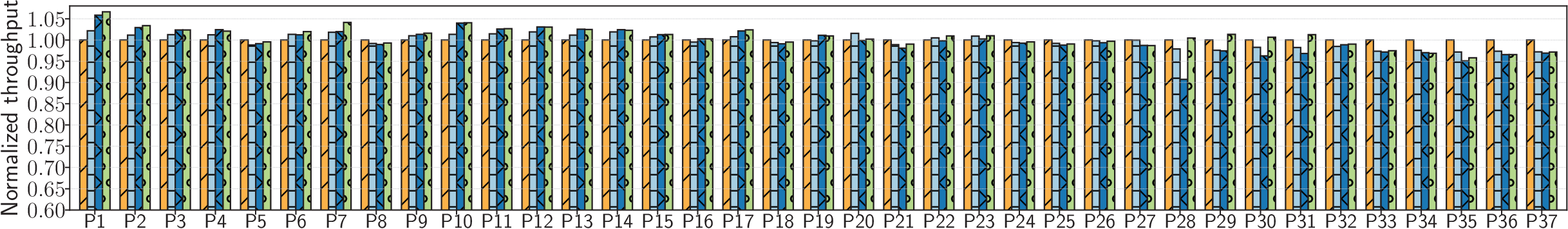}
\vspace{-0.6cm}
\caption{Normalized unfairness and STP values delivered by the dynamic cache-clustering policies for P$i$ workloads\label{fig:dynamic2}}
\vspace{-0.3cm}
\end{figure*}

Overall our results reveal that LFOC and LFOC+ exhibit a fairer behavior than the other schemes for the vast majority of the workloads. Table~\ref{table:mean} showcases the average and maximum reduction in unfairness achieved by the various approaches relative to Stock-Linux. In particular, LFOC+ achieves up to 25.15\% fairness improvement, and performs in a close range (3.3\% on average) of the \textit{Best Static} approach. While LFOC+ only achieves a 3.92\% average fairness improvement w.r.t. LFOC, the differences between both approaches become especially pronounced for workloads featuring 4 or more cache-sensitive programs. For this kind of workloads (e.g., S5-S9 or S23-S27), LFOC+ reduces unfairness by up to 9.8\% over LFOC. That is due to the fact that LFOC assigns cache-sensitive programs to separate and small ($\leq$ 2-way) partitions, while LFOC+ effectively combines some of these applications in the same partitions. In addition, confining all streaming programs in a single partition enables LFOC+ to bring additional improvements. Regarding throughput, there is no clear winner: all the algorithms perform in a tight 3\% range, and their average throughput improvement is very small (between 0.16\% --KPart-- and 0.61\% --LFOC+--). Notably, as the number of memory-intensive applications in the workload increases, LFOC and LFOC+ begin to slightly degrade throughput over Linux. We elaborate on this aspect in the next section.

\vspace{1pt}
\textit{B) Study of the Dynamic Policies.} We now focus on the analysis of the dynamic version of the LFOC, LFOC+, Dunn and KPart policies, which rely on on-line performance monitoring. We leave for future work the evaluation of the dynamic version of CPA and the adaptation of its implementation to our experimental platform, which, as pointed out by CPA's authors~\cite{cpa-tpds20}, requires to run thousands of experiments to determine the threshold values used for online application classification. Nevertheless, as shown in Sec. 5.1.A, CPA's cache-clustering algorithm provides small fairness improvements especially for workloads including more than 3 critical applications or at least 3 aggressor programs; this is the case of most of the workloads we used.

For our analysis, we created a user-level implementation of Dunn, as this policy was originally proposed as a user-level strategy~\cite{selfa-pact17}. We also tested with the publicly available user-level C++ implementation of 
KPart-Dynaway~\cite{kpart-github} --the dynamic version of KPart. Unfortunately, on our platform KPart-Dynaway's execution  crashes shortly after the partitioning algorithm is executed for the first time; this is due to a number of assumptions of this implementation that do not apply to our experimental setting~\cite{lfoc}. While this issue prevented us from evaluating KPart-Dynaway's on our platform, we were still able to measure the completion time of the partitioning algorithm for workloads with less than 12 applications. For 11-application workloads KPart's algorithm takes 4.14ms on average. By contrast, LFOC's and LFOC+'s algorithms take 0.94us and 6.06us on average for 12-application workloads, respectively, and 0.96us and 6.16us for 16-application workloads. As we showed in Sec. 5.1.A, KPart fails to provide better fairness than LFOC's or LFOC+'s cache-clustering algorithms, whose completion times are several orders of magnitude shorter. 

In our OS-level implementations of LFOC and LFOC+, PMCs are sampled every 100M instructions during the fairness mode and every 10M instructions during the sampling mode. The partitioning algorithm for Dunn, LFOC and LFOC+ is executed every 500ms, as in previous work~\cite{selfa-pact17,lfoc}. To verify that the various dynamic approaches were working as expected, we first experimented with workloads S1-S27. The corresponding results --omitted due to space constraints-- exhibit very similar trends to those of Sec. 5.1.A, where static cache-partitioning was used.

Fig. \ref{fig:dynamic2} shows the normalized unfairness and throughput values delivered by the dynamic versions of Dunn, LFOC and LFOC+ for $Pi$ workloads. These workloads consist of 8, 12, 16 or 20 applications each, and include benchmarks that exhibit distinct long-term program phases with varying degree of memory intensity, such as \texttt{xz}, \texttt{astar}, \texttt{mcf}, \texttt{xalancbmk} or \texttt{blender}. Some of these applications go through highly cache-sensitive phases, so Stock-Linux delivers higher unfairness values in these scenarios. This is why all strategies achieve higher unfairness reductions w.r.t. to Linux, than for $Si$ workloads, as summarized in Table~\ref{table:meanv}. Yet, LFOC+ is capable to improve fairness over Dunn across the board (by up to 22.18\% for P28, and by 9.13\% on average). Moreover, LFOC+ still outperforms LFOC in term of fairness (4.94\% on average) while providing better system throughput for most workloads. Still, the average throughput improvement achieved by any of the algorithms relative to Stock-Linux is almost negligible (less than 0.66\%).

\begin{table}
\parbox{.35\linewidth}{
\caption{Average and max. unfairness reduction of the various strategies relative to Stock-Linux for experiments of Fig.~\ref{fig:dynamic2}\label{table:meanv}}
\vspace{-0.2cm}
\centering
\footnotesize
\fontsize{8.5}{8.5}\selectfont 
\addtolength{\tabcolsep}{-3pt}  
\begin{tikzpicture}[scale=0.94,transform shape,inner sep=0pt]
\node (table1) {
\begin{tabular}{lrr}
\toprule
  \textbf{Strategy} &      \textbf{Avg.} &      \textbf{Max.} \\
\midrule
Dunn        &   6.85\% &  19.79\% \\
LFOC        &  10.88\% &  33.52\% \\
LFOC+       &  15.27\% &  36.77\% \\
\bottomrule
\end{tabular}};
\end{tikzpicture}
}
\hfill
\parbox{.60\linewidth}{
\caption{Sampling mode statistics\label{table:stats}}
\vspace{-0.2cm}
{\fontsize{8.5}{8.25}\selectfont
\addtolength{\tabcolsep}{-3pt}   
\begin{tabular}{lrrr}
\toprule
Work. &  Fraction & Activations&  Avg. \\
&  			total time & per second & Time\\
\midrule
P28      &           0.438\% &                   0.556 &               6.295ms \\
P29      &           0.266\% &                   0.585 &               4.552ms \\
P30      &           0.330\% &                   0.601 &               5.482ms \\
P31      &           0.370\% &                   0.570 &               6.490ms \\
P32      &           0.223\% &                   0.580 &               3.850ms \\
P33      &           0.423\% &                   0.607 &               6.973ms \\
P34      &           0.363\% &                   0.543 &               6.690ms \\
P35      &           0.381\% &                   0.596 &               6.380ms \\
P36      &           0.421\% &                   0.559 &               5.435ms \\
P37      &           0.428\% &                   0.601 &               7.118ms \\
\bottomrule
\end{tabular}
}

}
\end{table}

To conclude the discussion of these experiments, several key observations are in order. First, as the number of cache-sensitive applications in the workload gets closer to the number of LLC ways (11), there is smaller room for improvement; we could corroborate this trend via extensive analysis with the PBBCache simulator. Second, LFOC and LFOC+ tend to even out the applications' performance degradation by substantially reducing the relative slowdown of cache-sensitive programs, at the expense of potentially increasing the slowdown of aggressor streaming programs. Sometimes this approach also results into throughput gains relative to Stock-Linux. Third, these throughput gains are usually smaller as we increase the number of applications in the workload. Intuitively, this has to do with the cumulative nature of the STP metric (see Eq.~\ref{eq:stp}) and the fact that cache-clustering algorithms have little effect in the slowdown of certain programs (e.g., light-sharing ones); for these programs a fixed contribution to the summation of Eq.~\ref{eq:stp} is applied to all strategies. Related to this observation, and by zooming in on the bottom chart of Fig.~\ref{fig:dynamic2} we can see that the modest relative STP improvements of the evaluated strategies fade as we go towards the rightmost part of the chart. Moreover, fairness-aware approaches begin to introduce slight throughput degradation w.r.t. Stock-Linux for some 20-application workloads (P28-P37). This occurs due to two interrelated factors: (1) memory bandwidth contention, which begins to be substantial as we increase the number of co-running memory-intensive programs, and (2) the fact that fairness-aware approaches tend to grant fewer LLC ways to streaming programs, whose performance is very sensitive to bandwidth contention. Also of special attention is the striking difference in the throughput figures of LFOC and LFOC+ for P28-P31. In this context, LFOC uses two 1-way partitions to confine streaming programs most of the time. As discussed in Sec.~\ref{sec:other-features}, this degrades the performance of streaming programs due to the competition for a single LLC way, which causes up to a 9.6\% throughput degradation (for P28). LFOC+, by contrast, can deliver acceptable throughput in this scenario by employing a single 5MB (2-way) partition to confine all streaming programs. In addition, the more advanced mechanism used by LFOC+ to distribute LLC-space between sensitive applications allows it to reduce unfairness by up to 20.6\% w.r.t. LFOC (for P28).

\begin{figure*}
 \begin{minipage}[tbp]{1\textwidth}
  \begin{minipage}[tbp]{0.66\textwidth}
  \vspace{-0.3cm}
  \centering
 \includegraphics[width=1\textwidth]{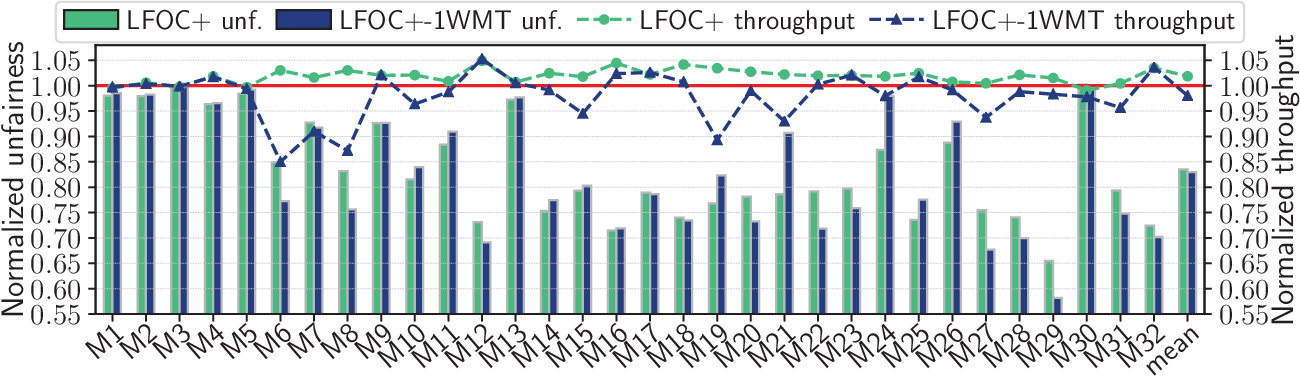}
\vspace{-0.7cm}
 \captionof{figure}{Unfairness and throughput provided by LFOC+ for workloads in Table~\ref{table:mt}\label{fig:multithreaded}}
  \end{minipage}
\begin{minipage}[tbp]{0.39\textwidth}
  \fontsize{8.5}{8}\selectfont
  \addtolength{\tabcolsep}{-3pt}    
 \captionof{table}{Multithreaded benchmarks considered\label{table:suites}}
 \vspace{-0.2cm}	
\begin{tabular}{ll}
\toprule
           Suite &                                              Benchmarks \\
\midrule
    PARSEC 3 &       blackscholes \seclass, bodytrack \seclass, \\
    		 & 		  fluidanimate \lsclass \\
	SPEC CPU2017 &  cam4\_s \seclass, lbm\_s \stclass, nab\_s \lsclass, \\
			& pop2\_s \stclass, xz\_s \seclass \\
    SPEC OMP2012 &                                   smithwa \stclass, swim \stclass \\
    Rodinia &         CFD-Euler3D \seclass, myocyte \lsclass, \\
    		& streamcluster \stclass \\
 	NAS Parallel B. &   BT \seclass, CG \seclass, EP \lsclass,  LU \seclass, SP \stclass \\
\bottomrule
\end{tabular}
    \end{minipage}
  \end{minipage}
\end{figure*}

\begin{table*}[tbp]
\vspace{0.1cm}
\caption{Workloads including multithreaded applications. The number of threads they run with are indicated in parentheses\label{table:mt}}
\vspace{-0.2cm}
\centering
\fontsize{8}{8}\selectfont
\addtolength{\tabcolsep}{-4pt}  
\begin{tikzpicture}[scale=0.94,transform shape,inner sep=0pt]
\node (table1) {\begin{tabular}{ll|} \hline
\toprule
Workload &                                                      Composition \\
\midrule
   M1 &                                                  bodytrack(10), BT(10) \\
   M2 &                                           cam4\_s(10),streamcluster(10) \\
   M3 &                                              bodytrack(10), pop2\_s(10) \\
   M4 &                                             bodytrack(10), smithwa(10) \\
   M5 &                                                  lbm\_s(10), cam4\_s(10) \\
   M6 &                                              CFD-Euler3D(10), swim(10) \\
   M7 &                                                    lbm\_s(10), xz\_s(10) \\
   M8 &                                             CFD-Euler3D(10), lbm\_s(10) \\
   M9 &                              nab17,gobmk06,povray06,xalancbmk06,LU(16) \\
  M10 &               deepsjeng17, fotonik3d17, omnetpp17, perlbench17, EP(16) \\
  M11 &                 lbm06, xalancbmk06, namd17, povray17, fluidanimate(16) \\
  M12 &                     gamess06, gromacs06, lbm06, omnetpp06, myocyte(16) \\
  M13 &        fotonik3d17, imagick17, povray17, xalancbmk17, blackscholes(16) \\
  M14 &                   GemsFDTD06, povray06, xalancbmk06, imagick17, BT(16) \\
  M15 &                  fotonik3d17, namd17, omnetpp17, x264ref17, cam4\_s(16) \\
  M16 &          GemsFDTD06, gamess06, omnetpp06, x264ref17, streamcluster(16) \\
\bottomrule
\end{tabular}};
\node [right=1pt of table1.north east,below right](table2) {\begin{tabular}{ll} \hline
\toprule
Workload &                                                      Composition \\
\midrule
 M17 &                     gamess06, povray06, xalancbmk06, imagick17, BT(16) \\
  M18 &               calculix06, omnetpp06, bwaves17, lbm17, fluidanimate(16) \\
  M19 &              GemsFDTD06, dealii06, wrf06, omnetpp17, streamcluster(16) \\
  M20 &               calculix06, lbm06, omnetpp06, bwaves17, fluidanimate(16) \\
  M21 &        gromacs06, namd06, lbm17, omnetpp17, fluidanimate(8), pop2\_s(8) \\
  M22 &           perlbench06, exchange217, namd17, omnetpp17, EP(8), lbm\_s(8) \\
  M23 &         h264ref06, omnetpp06, fotonik3d17, namd17, myocyte(8), xz\_s(8) \\
  M24 &                 gamess06, nab17, povray17, xalancbmk17, EP(8), swim(8) \\
  M25 &                 gobmk06, povray06, lbm17, omnetpp17, CG(8), myocyte(8) \\
  M26 &          bzip206, tonto06, xalancbmk06, roms17, blackscholes(8), CG(8) \\
  M27 &              h264ref06, lbm06, omnetpp17, perlbench17, EP(8), lbm\_s(8) \\
  M28 &               omnetpp06, perlbench06, nab17, wrf17, SP(8), nab\_s(8) \\
  M29 &           h264ref06, omnetpp06, zeusmp06, perlbench17, EP(8), lbm\_s(8) \\
  M30 &                  milc06, bwaves17, gcc17, xalancbmk17, EP(8), nab\_s(8) \\
  M31 &             bwaves06, dealii06, omnetpp17, parest17, SP(8), myocyte(8) \\
  M32 &  omnetpp06, perlbench06, wrf06, zeusmp06, myocyte(8), streamcluster(8) \\
\bottomrule
\end{tabular}};
\end{tikzpicture}
\vspace{-0.8cm}
\end{table*}

\vspace{-0.2cm}
\subsection{Overhead Analysis}\label{sec:overhead}

We now analyze the overhead of LFOC+'s sampling mode. To gather the required data from our OS-level implementation we used the SystemTap~\cite{systemtap} tool. For our analysis we employed the P28-P37 workloads; for these 20-application workloads the partitioning algorithm does not take long to run (7.17us on average), but the sampling mode is engaged more often than for other workloads, as they include many programs with a time-varying degree of cache sensitivity. Table~\ref{table:stats} shows the fraction over the total completion time that the sampling period is active under LFOC+, the frequency of activations of this mode, and the time required for applications in each workload to go through a sampling cycle. The results indicate that the sampling mode was active between 0.22\% and 0.44\% of the total execution time, and applications take 5.92ms on average to complete a sampling cycle. Recall that, unlike KPart, LFOC and LFOC+ do not always require a full cache-way sweep, and in most cases the sampling period can be interrupted as soon as the application class is identified. Notably, we observed that 75\% of the sampling mode activations take less than 11.5ms.  All in all, our experiments in Sec. 5.1.B reveal that the small percentage over the total execution time (0.36\% on average) devoted to sampling-mode processing enables LFOC+ to deliver substantial fairness improvements.

As stated earlier, we use a shorter instruction window for the sampling mode (10M instr.) than for the fairness mode (100M instr.). This enables to reduce the time required to complete sampling at the cost of extra overhead due to more frequent OS activations to process PMCs. To determine an upper bound of the overhead that comes exclusively from using the shorter 10M window, we ran all SPEC CPU programs alone on the system with both instruction windows, and found that using 10M windows degrades performance by up to 0.73\% (0.54\% on average) for the full execution.

\vspace{-0.2cm}
\subsection{Workloads Including Multithreaded Applications}\label{sec:multithread}

In this section we assess the effectiveness of LFOC+'s support for cache partitioning under workloads including multithreaded programs. Because none of the other partitioning schemes considered~\cite{selfa-pact17,kpart,lfoc,cpa-tpds20} support multithreaded applications, here we only provide the comparison between LFOC+ and Stock-Linux. Prior to our evaluation we performed an extensive analysis on the cache-sensitivity and contentiousness of the applications in four parallel benchmark suites: PARSEC3, SPEC OMP2012, NAS Parallel Bechmarks and Rodinia. We also experimented with the OpenMP implementation available for some SPEC CPU2017 FP \textit{speed} benchmarks. We found that many of the parallel benchmarks explored have a bandwidth-intensive cache-insensitive behavior, and their bandwidth consumption is very superior to the one observed in single-threaded programs from SPEC CPU. For example, in running a multithreaded application alone we observed that its bandwidth consumption can rise up to 66GB/s while for a single-threaded program it is usually no greater than 13GB/s. As a result, when using multithreaded programs, bandwidth contention has a great impact in throughput and fairness.

To build diverse multi-application workloads consisting of programs of different cache-sensitivity classes we picked a subset of multithreaded benchmarks from each of the five suites explored. Table~\ref{table:suites} enumerates the parallel benchmarks present in our multiapplication workloads, along with their cache-sensitivity class (i.e., \lsclass~light sharing, \seclass~sensitive, and \stclass~streaming). The selected benchmarks from PARSEC are POSIX threads programs, whereas all the others are OpenMP applications. Table~\ref{table:mt} shows the composition of the randomly-generated workloads explored in this section, which combine single-threaded and parallel benchmarks from different categories. Specifically, three types of workloads were considered: workloads M1-M8 are made up of 2 multithreaded programs, M9-M20 combine 4 singlethreaded and 1 multithreaded program, and M21-M32 consist of 4 single-threaded and 2 multithreaded programs. In all workloads the total number of threads matches the number of cores on our platform (20).

Fig.~\ref{fig:multithreaded} shows the unfairness and throughput delivered by LFOC+ relative to Stock-Linux. As a reference, we also display the results obtained when using 2.5MB 1-way partitions to confine multithreaded streaming programs instead of 5MB 2-way partitions (as LFOC+ does by default). This approach that treats all streaming programs similarly is referred to as \textit{LFOC+-1WMT}. We observe that LFOC+ provides substantial unfairness reductions when multithreaded programs are present in the workload. These gains (up to 34.5\% for M29) are especially pronounced for workloads that combine multithreaded and single-threaded applications. Clearly, LFOC+'s control to reduce bandwidth contention (i.e., using a 5MB partition to confine streaming multithreaded programs) constitutes an effective measure to reduce throughput degradation. Removing that control measure (i.e., \textit{LFOC+-1WMT}) may help reducing unfairness further, such as in M6, M8 or M20. Nevertheless, this may also backfire by causing noticeable throughput degradation (up to 15\% for M6). All in all, the bandwidth-contention control measure allows LFOC+ to make a more conservative cache-clustering approach when multithreaded applications are included in the workload, as it still offers substantial reductions in unfairness (16.5\% on average), with almost no throughput degradation across the board.

%% file: conclusions.tex
\vspace{-0.3cm}
\section{Conclusions and future work}\label{sec:conclusions}

In this article we have presented LFOC+, a dynamic OS-level cache-partitioning policy implemented in the Linux kernel. LFOC+ is a novel fairness-aware strategy that constitutes a substantial enhancement of our earlier LFOC policy~\cite{lfoc}. Our detailed simulation-based analysis of the fairness-wise optimal cache-clustering solution enabled us to guide the design of a more effective cache-clustering algorithm, used by LFOC+. This new  partitioning algorithm makes it possible to reduce the slowdown of cache-sensitive applications further 
by combining up to 2 of these applications in the same partition. Our experiments, conducted on an Intel CMP platform with hardware cache-partitioning support, reveal that LFOC+ improves fairness substantially over state-of-the-art partitioning policies, such as Dunn~\cite{selfa-pact17} (by up to 22.1\%), KPart~\cite{kpart} (up to 17.4\%), CPA~\cite{cpa-tpds20}~(up to 25\%) and LFOC~\cite{lfoc} (up to 20.6\%). The source code of LFOC+'s partitioning algorithm and that of our kernel-level partitioning framework has been released as part of PMCTrack v2.0 at \url{https://github.com/jcsaezal/pmctrack}. 

LFOC+ also includes specific support to effectively deal with data-parallel multithreaded applications, which are not handled by any other recent proposal~\cite{selfa-pact17,kpart,lfoc,cpa-tpds20}. We should highlight that this support may not be effective (1) when various threads of the same application exhibit very different cache-related behaviors, and (2) when a multithreaded program runs alone on the system, as LFOC+ does not partition the LLC in this case. To better deal with these scenarios, we strongly believe that cache-partitioning among threads could be exploited along with our inter-application partitioning approach. While some aspects of LFOC+, such as the PMC-based classification, may be also beneficial to guide intra-application partitioning (e.g., to isolate streaming threads from cache-sensitive ones), novel techniques still need to be devised to effectively support other types of multithreaded programs, such as latency-sensitive or throughput-oriented applications. For these programs, specific user-level metrics should be considered (e.g., jobs completed per unit time, latency, etc.) to measure the slowdown or to properly assess the degree of quality of service, along with potential cooperation from other software layers, such as the runtime system. As for future work, we also plan to design cache-clustering approaches for the latest AMD high-performance processors, where multiple logically independent LLCs --each one shared by a different subset of cores-- are present on the same chip~\cite{amd-ccx}. Although contention-aware thread-to-LLC mapping schemes have been proposed~\cite{sergeyb-edp,survey-contention,camps}, no previous proposal has yet addressed fair cache-clustering coupled with effective thread-to-LLC mapping. Optimizing fairness on these AMD processors is a more complex problem than on single-LLC CMPs, so this constitutes an interesting research avenue.